\documentclass[conference]{IEEEtran}
\IEEEoverridecommandlockouts
\usepackage{amsmath,amssymb,amsfonts}
\usepackage{algorithmic}
\usepackage{graphicx}
\usepackage{textcomp}
\usepackage[dvipsnames,table]{xcolor}
\def\BibTeX{{\rm B\kern-.05em{\sc i\kern-.025em b}\kern-.08em
    T\kern-.1667em\lower.7ex\hbox{E}\kern-.125emX}}
\usepackage[numbers]{natbib}
\usepackage{algorithm}
\usepackage{array}
\usepackage{amsthm}
\usepackage{multirow}
\usepackage{diagbox}
\usepackage{verbatim}
\usepackage[colorinlistoftodos]{todonotes}
\usepackage{amsopn}
\usepackage[font=small,labelfont=bf]{caption}
\usepackage{enumitem}%
\usepackage{float}%
\usepackage{subcaption}
\usepackage{amssymb}
\usepackage{tikz}
\usetikzlibrary{arrows}
\usepackage{hyperref}
\usepackage[capitalise]{cleveref}

\DeclareMathOperator*{\argmin}{arg\,min}

\DeclareMathOperator{\bigO}{\mathcal{O}}

\DeclareMathOperator{\action}{\mathbf{a}}
\DeclareMathOperator{\statee}{\mathbf{s}}

\DeclareMathOperator{\E}{\mathbb{E}} %

\usepackage{pifont} 
\newcommand{\cmark}{\ding{51}}%
\newcommand{\xmark}{\ding{55}}%
\newcommand{\linebreakand}{%
  \end{@IEEEauthorhalign}
  \hfill\mbox{}\par
  \mbox{}\hfill\begin{@IEEEauthorhalign}
}

\begin{document}

\title{Implementing Reinforcement Learning Datacenter Congestion Control in NVIDIA NICs}


\author{\IEEEauthorblockN{Benjamin Fuhrer\IEEEauthorrefmark{1}, Yuval Shpigelman\IEEEauthorrefmark{1}, Chen Tessler\IEEEauthorrefmark{2}, Shie Mannor\IEEEauthorrefmark{2}\IEEEauthorrefmark{3}, Gal Chechik\IEEEauthorrefmark{2}\IEEEauthorrefmark{4}, Eitan Zahavi\IEEEauthorrefmark{1}, Gal Dalal\IEEEauthorrefmark{2}}
\IEEEauthorblockA{\IEEEauthorrefmark{1}NVIDIA Networking, \IEEEauthorrefmark{2}NVIDIA Research, \IEEEauthorrefmark{3}Technion Institute of Technology, \IEEEauthorrefmark{4}Bar-Ilan University}}



\maketitle
\begin{abstract}
As communication protocols evolve, datacenter network utilization increases. As a result, congestion is more frequent, causing higher latency and packet loss. Combined with the increasing complexity of workloads, manual design of congestion control (CC) algorithms becomes extremely difficult. This calls for the development of AI approaches to replace the human effort. 
Unfortunately, it is currently not possible to deploy AI models on network devices due to their limited computational capabilities. 
Here, we offer a solution to this problem by building a computationally-light solution based on a recent reinforcement learning CC algorithm \cite[RL-CC]{rl-cc}.
We reduce the inference time of RL-CC by x500 by distilling its complex neural network into decision trees. This transformation enables real-time inference within the $\mu$-sec decision-time requirement, with a negligible effect on quality. We deploy the transformed policy on NVIDIA NICs in a live cluster. Compared to popular CC algorithms used in production, RL-CC is the only method that performs well on all benchmarks tested over a large range of number of flows. It balances multiple metrics simultaneously: bandwidth, latency, and packet drops. 
These results suggest that data-driven methods for CC are feasible, challenging the prior belief that handcrafted heuristics are necessary to achieve optimal performance.
\end{abstract}

  
  


\begin{figure*}
    \centering
    \includegraphics[width=\textwidth]{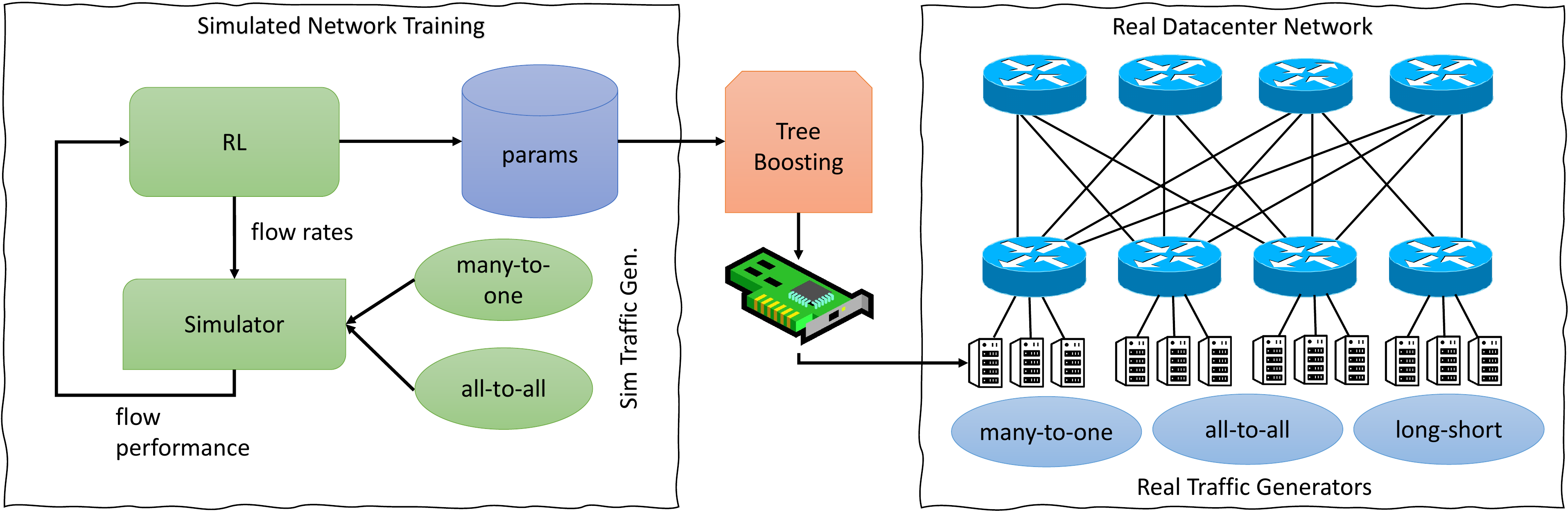}
    \caption{An overview of the deployment process of RL-CC (reinforcement learning congestion control) in the real world. From left to right: (1) an RL policy is trained in simulation; (2) the neural network policy is distilled into a compute and memory efficient tree-based representation; and (3) the tree policy is deploy on ConnectX-6Dx NIC firmware and tested in a live datacenter with standard benchmark traffic patterns.}
    \label{fig: visualization}
\end{figure*}

\section{Introduction}

Modern datacenters support computationally intensive applications such as distributed data processing, heterogeneous and edge computing, and storage. With advances in hardware and software, networks can support bandwidths up to 400Gbps (e.g., NVIDIA ConnectX-7 \cite{burstein2021nvidia}). At such speeds, the typical remote memory access, traditionally handled by the remote CPU, becomes a bottleneck. CPU over-utilization also leads to application delays and an increase in operational costs.
A natural solution is to offload memory management to the network interface card (NIC). Remote Direct Memory Access (RDMA) and RDMA over converged Ethernet \cite[RoCEv2]{beck2011performance} provide protocols that bypass the CPU, resulting in lower CPU overhead and higher bandwidth.
Consequently, RDMA has been increasingly adopted in datacenter networks \cite{rdma_widespread_use}.

Therefore, traffic congestion becomes the limiting factor in network performance.
Congestion occurs when traffic arrives at a node---switch or NIC---at a faster rate than it can be processed. Since each node is equipped with a first-in-first-out queue, the transmission latency increases monotonically with congestion. Therefore, efficient congestion control (CC) is crucial to maintaining high throughput and low latency in datacenters. CC algorithms limit the transmission rate or the number of bytes in the network of each flow (connection). By observing changes in the network, such as latency and telemetry signals, these algorithms are tasked with preventing congestion. Successful CC algorithms should react to network changes. Such changes occur on the order of $\mu$ seconds as a result of the reduction in latency achieved by RDMA.

This low-latency limitation fits well with heuristic-based CC algorithms. Because they rely on pre-defined rules, they can make decisions rapidly on low-compute devices. Due to their handcrafted nature, these methods tend to perform exceptionally well in a specific set of tasks. However, they tend to underperform in others for which they were not optimized. For example, DCQCN \cite{DCQCN} and Swift \cite{SWIFT} have been optimized for steady-state scenarios. But, as shown in \cite{rl-cc}, their reaction time is too slow to respond to sudden bursts of short flows.

Recently, \citet{rl-cc} introduced a data-driven approach that automatically learns a CC policy. They devised an algorithm that optimizes latency and bandwidth throughout multiple steps. Their method resulted in a robust policy capable of handling a range of tasks in a \emph{simulated} network.
Despite its impressive results, the method in \cite{rl-cc} has a major flaw that hinders its applicability: it relies on a neural network (NN) architecture. Inference using deep NNs requires massively parallelized computations. Unfortunately, such capabilities are out of reach for present networking devices. Even when advanced quantization and pruning techniques are used, the inference time remains too slow for the CC algorithm to react to network changes and successfully prevent congestion.

\footnote{The source code and simulator used to train RL-CC are available at: \href{https://github.com/NVlabs/RLCC}{https://github.com/NVlabs/RLCC}.}


In this work, we overcome the above issues that prevented ML-based CC approaches from reaching production pipelines. 
Our main contributions are:
\begin{enumerate}

    \item We show how to map complex policies to a computationally light architecture, gaining x500 inference-time reduction with a negligible effect on policy quality. Specifically, we map a deep NN to decision trees and reduce the inference time from $450~\mu sec$ to $0.9~\mu sec$.
    

    \item Leveraging NVIDIA's programmable congestion control \cite{pcc}, we deploy our method on production ConnectX-6Dx NICs in a cluster with 64 hosts over RoCE lossy fabric. We achieve state-of-the-art results in extensive evaluations that outperform DCQCN and Swift.

    \item Finally, we analyze the decision-making process of RL-CC, showing that it has learned non-trivial behaviors depending on past and current state. These results can be of independent interest to the general CC community, aiding in the design of future CC algorithms.
\end{enumerate}

Our RL-CC training and production pipeline is visualized in Fig.~\ref{fig: visualization}.

\section{Background and Problem Setup}
In this section, we provide an overview of the relevant background and prior work, and formulate the problem setup.
\subsection{Congestion Control}\label{background: cc}
RoCEv2 can be implemented in lossless and lossy networks \cite{rocerocks}. In lossless networks, Priority Flow Control (PFC) prevents packet drops by suspending transmission. This behavior has been shown to cause congestion spread that may impact other flow performance. It may also cause other problems, such as deadlock \cite{DCQCN,pfc_deadlock}. On the other hand, in lossy networks, dropped packets are re-transmitted. This results in increased latency and a reduction of goodput---net bandwidth, excluding re-transmissions. CC algorithms have been demonstrated to minimize PFC activation in lossless networks and reduce packet drops in lossy networks, thus improving overall network performance \cite{RoCC,DCQCN,TIMELY,rocerocks,SWIFT,cong_in_lossless,breaking_one_rtt,fasttune}. These methods govern the transmission rate of each flow while balancing multiple, possibly conflicting, objectives. They aim to maximize network utilization and fairness between flows while minimizing packet latency and packet drops.

The conflict between these objectives was explained by \citet{SWIFT}: due to the statistical nature of real systems, when $N$ flows share a congested path, and each transmits at the optimal rate ($\text{line rate}/N$), the average queue length is $\bigO(\sqrt{N})$. Hence, any low latency or high bandwidth solution that results in a different buffer occupancy results in a transmission rate trade-off between flows.

Although the above trade-off is clear when in a steady state, CC algorithms must also adapt rapidly to changes and converge to a new equilibrium. This can happen when flows abruptly stop transmitting, or alternatively when new ones join and start transmitting.
Maintaining a stable steady state and reacting quickly to changes are contradictory abilities. High sensitivity to changes in transmission rate impairs convergence to a stable point. Contrarily, small changes to the transmission rate may cause it to converge too slowly, resulting in packet drops or under-utilization.

\subsection{Existing State-of-the-Art for CC}\label{background: existing cc}
To evaluate the network's status and adjust the transmission rate appropriately, existing state-of-the-art (SOTA) CC relies on indications such as end-to-end delay and switch queue length. Those deployed in practice utilize rule-based heuristics to react to such indications. For example, DCQCN \cite{DCQCN}, a popular CC algorithm in datacenter deployments, utilizes Explicit Congestion Notification (ECN) \cite{ECN}. As ECN packets are statistical indications of developing congestion, DCQCN reduces the transmission rate once such packets are observed. Other algorithms such as Timely \cite{TIMELY} and Swift \cite{SWIFT} rely purely on end-to-end latency measurements for making decisions. Lastly, HPCC \cite{HPCC} takes as input the switch queue length and port bandwidth. This information, called telemetry, is only accessible in datacenter networks with appropriate hardware support. 

Varying traffic patterns and network topologies result in different indicator statistics. Thus, a common drawback of conventional CC algorithms is the need for manual tuning of their multiple parameters. Such tuning necessitates laborious calibrations by domain experts. And yet, the results are often unsatisfactory at properly balancing the tradeoffs (\cref{background: cc}). For example, DCQCN excels at stability in steady-state workloads such as storage but is slow to adapt to more dynamic compute-heavy workloads \cite{HPCC}. HPCC, on the other hand, is a top-performer in dynamic workloads at the expense of stability and high utilization during steady-state scenarios \cite{RoCC}.

In this work we evaluate CC on a live cluster. However, as our switches lack the appropriate telemetry support, our comparisons are limited to DCQCN and Swift. We refer the reader to \cite{rl-cc} for simulated comparisons with HPCC.

\subsection{Transmission Rate Modulation}
Typically, CC is conducted by setting a maximum transmission rate per flow (rate-limiting). 
Traditional transmission rate modulation uses Additive Increases and Multiplicative Decreases (AIMD) \cite{aimd_convergence}. \citet{aimd_convergence} showed that by performing a fixed additive increase while congestion is not observed and halving the transmission rate otherwise, AIMD converges to a fair solution where all flows utilize an equal share of the network. In addition, they argued that other additive/multiplicative variations, AIAD, MIAD, and MIMD, do not reach fair solutions.

With the emergence of high-speed links, the classic AIMD algorithm was shown to under-use link capacity \cite{highspeed_aimd_fail}. Since then, CC algorithms have evolved and more sophisticated methods have been proposed to modulate transmission rates. For instance, DCQCN increases rate by multiple successive increases towards a target rate, followed by slow increments of the target rate itself. Once congestion is observed, it reduces the transmission rate by $\alpha$, a parameter that the algorithm dynamically adjusts.
Similarly, Swift is also an AIMD variant. When receiving an ACK packet, Swift uses the difference between a flow delay and a target value to determine the rate change. In contrast, HPCC applies both AIMD and MIMD to avoid congestion. AIMD is used to maintain a stable steady state, whereas MIMD is used when changes in the network occur and a rapid reaction is necessary to recover bandwidth.

Due to its excellent stability property, the prior work described above and others mostly focused on AIMD. MIMD, on the other hand, is trickier; while it allows faster recovery and lower packet latency, it requires careful tuning to be able to reliably reach convergence. This is where artificial intelligence (AI) naturally fits to fulfill its promise of adaptively tuning its behavior within complex data patterns.

\subsection{Networking solutions based on AI}

Machine learning (ML) has been successfully applied in numerous disciplines, from healthcare to autonomous driving. Compared with manual tuning methods, ML algorithms can extract complex patterns from vast amounts of data and learn implicit correlations that enable better generalization and performance. Previous work considered ML-based solutions for networking problems \cite{ml_in_cc}. However, these algorithms require a lot of memory and are computationally demanding. Generally, for CC algorithms to operate successfully, their decision time must be $\bigO(\text{RTT})$. For modern datacenters that use RDMA, this is on the order of $1$ to $2\mu sec$. These limitations partially explain why there are currently no learning-based CC algorithms in production.

\citet{vivas} described PCC Vivace, an algorithm that performs online optimization over a utility function. As an online algorithm, its training and inference stages are interleaved when deployed. PCC Vivace runs gradient ascent, which is too computationally demanding for current NIC hardware. Their work was then expanded by \citet{jay2019deep}, which introduced Aurora, a CC framework based on deep RL. Aurora is designed for a single flow only; hence, it can only be applied to toy domains. \citet{sdn-rl} proposed two RL CC algorithms, one based on Q-learning \cite{qlearning} and the other on SARSA \cite{sarsa}. Unfortunately, \citet{sdn-rl} assume that the agent has joint access to all flows, which is not feasible in many networks, including those we study here. \citet{ndn-rl} developed an RL CC algorithm for Named Data Networking -- conceptual future networks that assume the agent has knowledge of the flow origin application. \citet{rl-satellites} applied an RL algorithm called DDPG \cite{ddpg}, to learn CC strategies in multipath TCP used in satellite networks.

Recently, \citet{rl-cc} introduced an RL-based RDMA CC algorithm called RL-CC. It is the first and only AI algorithm to successfully tackle multi-flow traffic scenarios relying only on RTT measurements. It thus fits most networks that exist today, and we choose to build upon it in this work. In several network simulation benchmarks, RL-CC outperformed SOTA rule-based CC algorithms: DCQCN, SWIFT, and HPCC. One key feature of RLCC is MIMD rate modulation as a function of historic observed congestion. Another reason for the success of RL-CC is its carefully designed reward function, which at its optimum embodies an optimal flow equilibrium. In line with our summary above, \citet{rl-cc} state that they would require dedicated hardware to accommodate the computational burden of deep learning inference.

In this work, we build on RL-CC. First, we analyze the parameters that affect the reward function. Then, we tackle the computational burden. Specifically, we show that due to the slow inference of NN architectures on the ConnectX-6Dx devices, RL-CC fails to operate successfully. We solve this issue with distillation techniques, mapping the learned policy to a lighter architecture of decision trees. Finally, we perform an extensive analysis of RL-CC evaluating it on large-scale tests on real hardware devices.

\section{RL for CC}

Congestion control is a sequential decision-making problem. The decision maker is an instance of the CC algorithm that runs within the NIC and controls the rate of a single transmission flow. From now on, we refer to the decision maker as an \emph{agent}. The agent acts on the latest information available to that instance, including the current and past transmission rate, the RTT, and the last actions taken. As the agent interacts with the network by sending an RTT packet and modulating its transmission rate, the agent cannot access information regarding other concurrent agents and their state. Therefore, the agent must act strictly on the basis of its local state. 

 


Formally, we model this task as a multi-agent partially observable Markov decision process (POMDP) \cite{rl-cc}. At each step, the agent is in some state of the POMDP and observes a corresponding partial observation. Based on the current observation, the agent chooses an action. This mapping from observations to actions is called a \emph{policy}. Once the agent acts, the agent transitions into a new state and receives a reward. Its goal is to obtain the highest average reward along the trajectory in expectation w.r.t. the stochasticity in the system. The agent achieves this goal by finding the best policy possible. We aim to devise a reward function that reflects a good balance between BW and latency with minimal packet loss by considering an agent's RTT and transmission rate. The outcome of our training process is the policy.



More explicitly, our POMDP consists of:

\textbf{Observations.} The agent observes information relevant only to the flow it controls. This includes the current and past transmission rates, the RTT measurement, and its previous decisions.

\textbf{Actions.} At time $t,$ the agent selects an action $\action_t$ that modifies the next transmission rate in a multiplicative manner, $\text{rate}_{t+1} = \action_t \cdot \text{rate}_t$. 


\textbf{Reward.} \citet{SWIFT} have shown that network congestion is optimized when all $N$ flows sharing a congested path are rate-limited to exactly their fair share of $\frac{\text{line rate}}{N}$ and that then the average queue length is $O(\sqrt{N})$.
We define $\text{RTT-inflation}=\frac{\text{RTT}}{\text{base-RTT}}$ as the RTT normalized by its measurement in an empty system. As the RTT grows monotonically with the transmission rate, their product is constant when the transmission rate is fixed. We define $\textbf{target}=\text{RTT-inflation}\cdot\sqrt{rate}$, where $rate = \frac{\text{line rate}}{N}$, i.e. flows transmit at the ideal rate. We call \textbf{target} the inflation control parameter as it tunes the expected steady-state RTT-inflation. 

We extend the reward from \cite{rl-cc}, that was inspired by \cite{SWIFT}, by adding a congestion tolerance parameter $\beta$. The role of $\beta$ is to avoid aggressively reducing the transmission rate when the buffer occupancy is low. We define the reward obtained by agent $i$ controlling flow $i$ at time $t$ as:
\begin{equation}
    r_t^i = -\left( \text{\bf{target}} - \text{max}(\text{RTT-inflation}^i_t - \beta, 0) \cdot \sqrt{\text{rate}^i_t} \right)^2 \,.
    \label{eq_reward}
\end{equation}
The reward has the benefit that the system achieves a fixed-point equilibrium when the reward is maximal i.e. $r=0$. The agent maximizes the reward by modulating the transmission rate to minimize the distance between the current RTT-inflation and its target value at steady-state.  

We clearly see that when ${\text{RTT-inflation} > \beta},$ the RTT-inflation in steady-state in expectation is expressed as:
\begin{equation}
    \E[\text{RTT-inflation}] = \text{\bf{target}}\cdot \sqrt{\frac{N}{\text{line rate}}} + \beta \,,
    \label{eq_rtt_inflation}
\end{equation}

In the following section, we provide an extensive analysis of how the user-chosen parameters $\textbf{target}$ and $\beta$ affect the agent's behavior. 



\textbf{Policy optimization:} The CC environment is particularly challenging compared to the standard MDPs that RL algorithms usually tackle. That is because of a unique combination of a partially observable multi-agent system with multiple objectives that are also non-stationary. For these reasons, \citet{rl-cc} developed the custom-made deterministic on-policy algorithm that leverages access to the analytical form of the reward function. The analytical form enables training an RL agent to solve the CC environment; hence the algorithm is called RL-CC. Plugging in our extended reward function from \eqref{eq_reward}, we obtain the following policy gradient approximation
\begin{align}
\label{eq: gradient term}
    \nabla_{\theta} \rho^{\pi_\theta} (\statee^i) &\approx \Bigg[  \lim_{T \rightarrow \infty} \frac{1}{T} \sum_{t=0}^{T} \Big(\text{\bf target} - \\ \nonumber
    &\text{max}(\text{RTT-inflation}^i_t - \beta, 0) \cdot \sqrt{\text{rate}^i_t}
    \Big) \Bigg] \nabla_\theta \pi_\theta (o(\statee^i)) \, ,
\end{align}
where $\theta$ are the policy parameters (NN weights), $ \rho^{\pi_\theta}$ is the average reward along a $T$-step trajectory, and $o(\statee^i)$ is the observation in state $\statee^i.$
For ease of notation, we define:
\begin{align}
\label{def: delta}
    \delta^i_t = \text{\bf target} - \text{max}(\text{RTT-inflation}^i_t - \beta, 0) \cdot \sqrt{\text{rate}^i_t}.
\end{align}
For states in which $\delta^i_t < 0,$ a negative weight is assigned to the policy gradient, effectively reducing the transmission rate in those states, and vice versa for $\delta^i_t > 0.$ Furthermore, $\delta^i_t \in [-\infty, \text{\bf target}]$ is bounded from above as the minimal RTT inflation is 0, but not from below as the latency can grow arbitrarily large. As a result, the policy is influenced to react more aggressively towards congestion.

\textbf{NN Architecture.} RL-CC's architecture \cite{rl-cc} was originally composed of two fully connected layers (input$\rightarrow$32$\rightarrow$16) followed by an LSTM layer \cite{LSTM} (16$\rightarrow$16) and then an output fully connected layer (16$\rightarrow$1). The input is the current state, $\statee_{t}:=(\delta_t, \action_{t-1})$, a tuple consisting of $\delta$ at the current timestep, and the action taken in the previous timestep. The LSTM hidden states are unique for each flow, enabling the policy to incorporate the flow's specific past information into the decision. Consequently, RL-CC is able to handle the partial observability of the environment. 

\begin{figure}
    \centering
    \includegraphics[width=\linewidth]{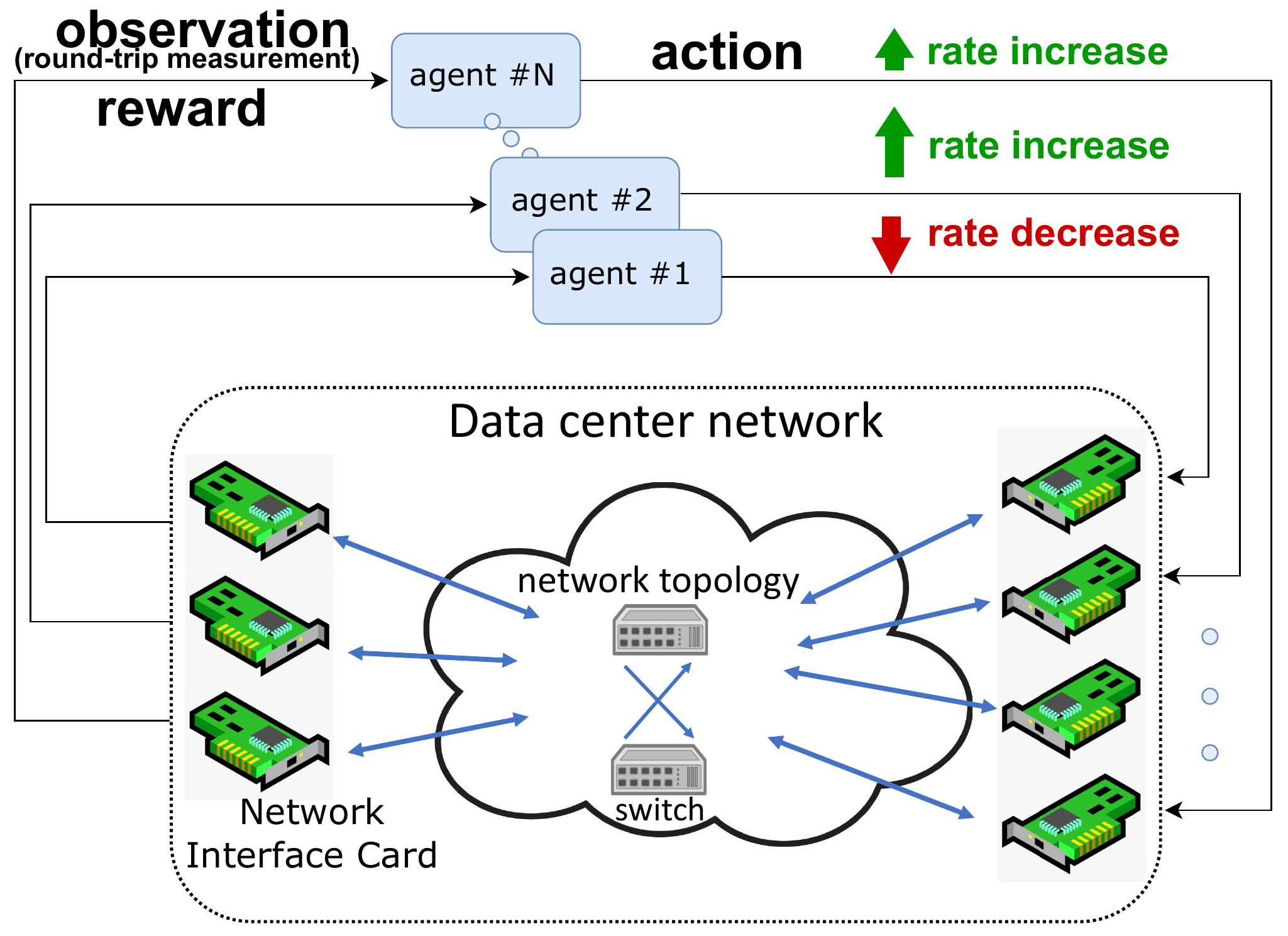}
    \caption{\textbf{RL-CC training loop.} Each flow is controlled by a different copy of the same agent, sharing the same logic across all flows but with its own local history. The agent interacts with the environment by multiplicative increment or decrement of the flow transmission rate (for visualization only we drew here a single flow per NIC). The environment feedback is the RTT measurement per flow.}
    \label{fig: rl-cc training loop}
\end{figure}

\textbf{Training.} Recall that each flow is controlled by a different copy of the same agent. The agent interacts with the environment by modulating the flow transmission rate. After each interaction, the local history of previous states, rewards, and actions is added to a fixed-sized rollout buffer. When the buffer is full, the policy gradient is calculated using \cref{eq: gradient term} and is used to update the policy. This procedure is repeated until the policy successfully maximizes the reward across flows. The RL training loop is visualized in \cref{fig: rl-cc training loop}.

In \cref{sec:analysis} we analyze various aspects of RL-CC in simulation. 
In \cref{sec:lightweight-rl} we distill the NN RL policy to light decision trees to enable deployment on a real device. Lastly, we test our distilled agent on a live cluster and explain its decisions on an example scenario.


\section{Design considerations for RL-CC}\label{sec:analysis}
Our goal is to deploy RL-CC in the real world. We begin with an in-depth analysis of various design decisions and how the RL agent can be controlled.

Here, we focus on simulation. While the simulations are rich, they do not precisely mimic real-world behavior. For example, they do not perfectly model the performance of the application software, host bottlenecks, and specific transport operations. Thus, later in Section~\ref{sec: experiements} we also present live experiments. For simulation, we use a realistic OMNeT++ emulator \cite{varga2002OMNeT++} that models NVIDIA ConnectX6-Dx NICs within a single-switch network. We experiment with different combinations of total flows, ranging from $2$ to $8192$, distributed across multiple hosts. We train the RL agent on a small set of benchmarks and then evaluate it on more complex ones, with precise in-network measurements. 

 We begin with a comparison of theory and practice. We show how well the performance observed in the simulation matches the theoretical expected performance. Then, we explain the role of the controllable parameters $\text{\bf target}$ and $\beta$, and analyze their effect on the behavior of RL-CC. We performed the simulation in this section on a many-to-one scenario (4 to 1) for one simulated second. Unless mentioned otherwise, we repeated the experiment with different numbers of flows per host and collected the data for a period of 0.5 sec after reaching a steady state. For our parameters, we used $\text{\bf target}=0.064,~ \beta=1.5$.

\begin{figure}
\centering
    \includegraphics[width=\linewidth]{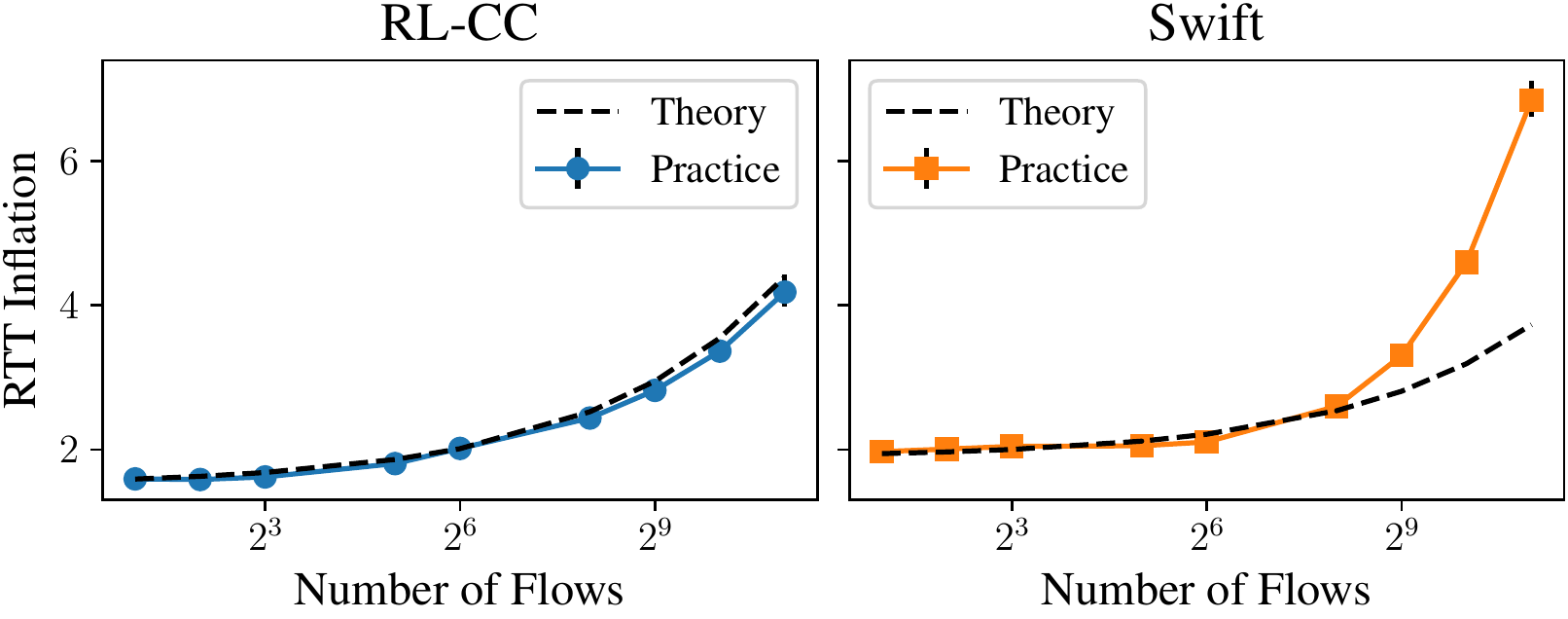}
    \caption{RL-CC and Swift Theory vs. Practice: RTT inflation as a function of number of flows. Curved lines represent theoretical curves in the order of $\bigO(\sqrt{N}).$ We plot the average RTT inflation per flow with 99\% vertical confidence intervals. Error bars are small initially and grow as the number of flows increase.} 
    \label{fig: sim vs theory rtt }
\end{figure}

\textbf{Theory versus practice:} In \cref{fig: sim vs theory rtt }, we compare the theoretical and practical RTT inflation for RL-CC and Swift. We calculated the theoretical values following \cref{eq_rtt_inflation} for RL-CC and use the best fitting $\bigO(\sqrt N)$ curve for Swift. As seen, RL-CC fits the theoretical curve perfectly. This confirms that the agent converges to an optimal policy that saturates the link while maintaining similar transmission rates amongst flows. Swift, on the other hand, diverges from the fitted curve as the number of flows increases. We attribute this behavior to Swift's AIMD rate modulation. The relative \textit{additive} rate adjustment is higher in percentage at low transmission rates. This becomes apparent with the increase in flows, further reducing the individual transmission rates. This combination results in overshoots, which negatively affect latency.


\textbf{Reward design:} RTT may increase even when the combined transmission rate of all flows is below the maximal rate. This increase happens due to stochastic collisions between flows \cite{SWIFT} and leads to an unnecessary decrease in transmission rate. The congestion tolerance parameter $\beta$ prevents this behavior by encouraging flows to increase their rate as long as the RTT is below $\beta$. As a result, the bandwidth increases when the number of flows is small. Furthermore, the significance of $\beta$ decreases as the number of flows increases, resulting in a minor impact on the delay when the number of flows is large. Once the RTT inflation exceeds $\beta$, the algorithm becomes sensitive to inflation indications. At this stage, $\text{\bf target}$ controls the bandwidth-latency trade-off. 

\begin{figure}
    \centering
    \begin{subfigure}[t]{1\linewidth}
    \centering
    \caption{Varying $\beta$}
        \includegraphics[width=0.75\linewidth]{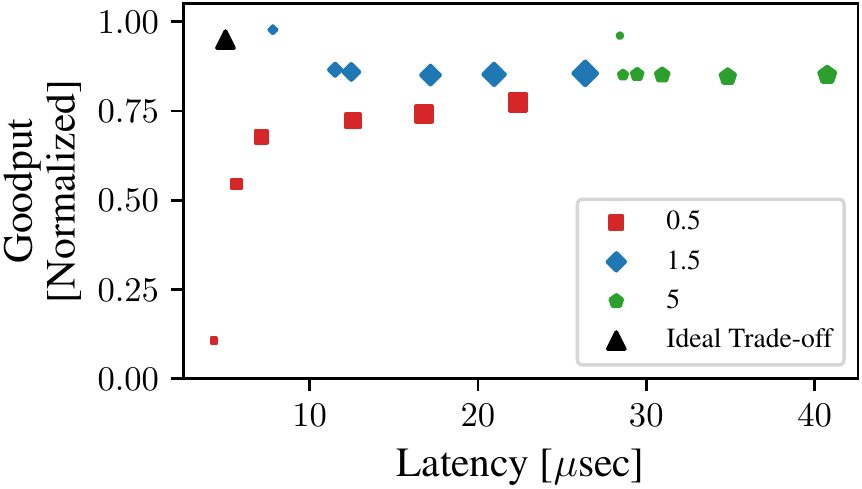} 
    \end{subfigure}
    \begin{subfigure}[t]{1\linewidth}
    \centering
    \caption{Varying target}
        \includegraphics[width=0.75\linewidth]{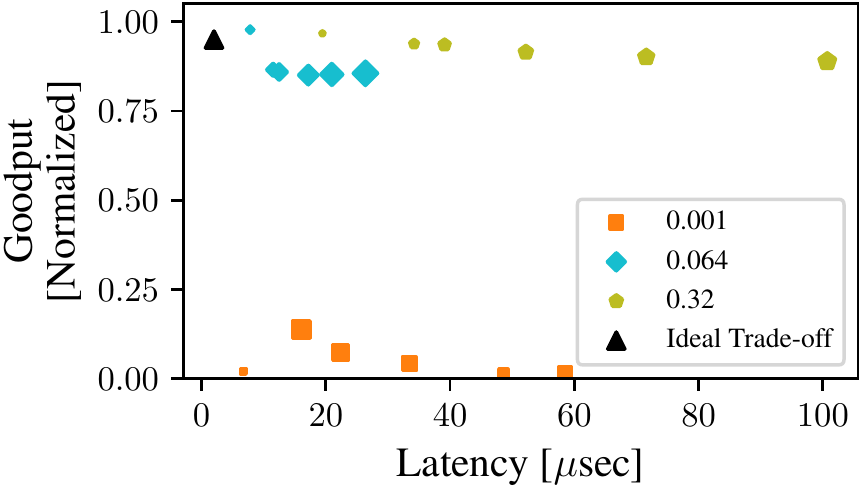} 
    \end{subfigure}
    \caption{RL-CC parameter influence on the bandwidth/latency tradeoff. The plot on the top presents the tradeoff when varying $\beta$, whereas on the bottom the effects of \textbf{target}. We observe that while lower beta correlates with lower latency, the agent fails when $\beta$ is set too low. On the other hand, when \textbf{target} was set too low, the agent fails yet a value too high results in a dramatic increase in latency. We found that the optimal values are $\beta=1.5$ and $\text{\textbf{target}}=0.064$. 
    } 
    \label{fig: beta_target_tradeoff }
\end{figure}

\begin{table*}
\parbox{.45\linewidth}{
\center
\begin{tabular}{c|c|c}
                     & \textbf{Flops} & \textbf{Decision Latency {[}$\mu sec${]}} \\ \hline
\textbf{LSTM}        & 2600           & 450                                      \\
\textbf{MLP}         & 200            & 17                                       \\
\textbf{Tree (ours)} & -              & \textbf{0.9}                                      \\ \hline
\end{tabular}

    \caption{\textbf{FLOPS and inference latency} as calculated on ConnectX-6Dx. We compare three architectures, LSTM (memory-based recurrent NN), MLP (fully connected), and our tree model. Binary decision trees can efficiently run directly on the NIC when implemented as a sequence of if-else statements. This significantly reduces the decision latency and enables our tree model to meet the required inference-time limit. 
    }
    \label{table: operations}
}
\hspace{0.8cm}
\parbox{.45\linewidth}{
    \begin{tabular}{c|cc|cc|cc}
        \textbf{Num. Flows} & \multicolumn{2}{c|}{\textbf{16}} & \multicolumn{2}{c|}{\textbf{64}} & \multicolumn{2}{c}{\textbf{128}} \\ \cline{1-1}
        \textbf{Metric}    & GP              & Drops          & GP            & Drops            & GP            & Drops            \\ \hline
        \textbf{LSTM}      & \textbf{0.91}           & 0              & 0.33         & 60K          & 0.40         & 16K         \\
        \textbf{MLP}       & \textbf{0.91}           & 0              & 0.31          & 60K         & 0.39         & 16K         \\
        \textbf{Tree (ours)}      & 0.86           & 0              & \textbf{0.86}         & \textbf{1.59}          & \textbf{0.85}          & \textbf{6.96}          \\ \hline
    \end{tabular}
    \caption{\textbf{Policy Distillation: Comparing MLP (fully-connected) based policy vs distilled policy} on many-to-one scenarios 
    while varying the number of flows. We measure the normalized goodput (GP), and average packet drops per flow on a \textbf{single switch cluster} equipped with 7 ConnectX-6Dx NICs. Although both MLP and tree performed similarly in simulation, when evaluating on real devices, the impact of the decision latency becomes apparent. Slow reaction time leads to high packet loss and a dramatic drop in goodput.}
    \label{table: mlp_vs_tree_deployed}
}
\end{table*}

\citet{SWIFT} 
have demonstrated that increasing the $\text{\bf target}$ value increases bandwidth at the expense of latency. Therefore, we aim to choose the lowest possible $\text{\bf target}$ to reduce latency while preserving competitive bandwidth. Furthermore, they showed that the bandwidth rapidly drops below a certain $\text{\bf target}$ value. The exact $\text{\bf target}$ lower bound is a function of the characteristics of the network. In \cref{fig: beta_target_tradeoff } we present an ablation study of several values, both for $\text{\bf target}$ and $\beta$. We observe that a low $\beta$ results in instability when the number of flows is small. This is due to the behavior of the reward function and the system's statistical nature, 
The plot also shows how a higher $\text{\bf target}$ leads to higher throughput, but also higher latency. On the other hand, when $\text{\bf target}$ is too small, the system is required to maintain a near-empty queue. This leads to unstable performance.

\section{Deploying RL-CC}\label{sec:lightweight-rl}

Previous sections covered RL-CC \cite{rl-cc} and analyzed various design decisions. Our experiments there, as well as those in \cite{rl-cc}, were carried out in \textit{simulation}. In this section, we present the challenges of deploying RL-CC in production given constraints on low memory and low inference time on limited hardware. We then show how to distill our RL agent to lightweight decision trees with negligible effects on performance, and conduct various experiments on a live cluster.

\subsection{Limitations of Neural Networks in NICs}\label{deployment_limitations}
As discussed above, RL-CC neural network based policy limits the flow rate to avoid congestion.
We begin by explaining why neural networks cannot be deployed on existing NICs. RDMA networks typically have low latency, with an RTT around $10 \mu sec$. Because RL-CC acts on the basis of RTT measurements, its inference-time needs to be significantly lower than that. As a result, we have set a decision-time upper bound of $\sim 2 \mu sec$. The decision-time measurement begins upon receiving an RTT packet and up until the algorithm modulates the transmission rate, which includes inference-time. We note, that the decision times for both DCQCN and Swift are below $2 \mu sec$.


NVIDIA's ConnectX-6Dx introduces a programmable CC engine that exposes an SDK for CC implementation. The code runs on in-data-path microprocessors that interact quickly with the NIC's send/receive pipes. This mechanism has a limited amount of global and per-flow memory. Furthermore, the processor's instruction set does not support floating-point operations or mathematical libraries for implementing deep-learning activation functions. RL-CC's original architecture 
is composed of two fully-connected layers 
followed by an LSTM layer \cite{LSTM} 
and then an output fully-connected layer.
  This architecture sums up to over $3000$ FLOPS, 2753 parameters (weights) stored in shared memory, and 32 flow-specific parameters (LSTM hidden states) stored in per-flow memory. Memory restrictions are $\bigO(\textbf{hundreds} \text{ of bytes})$; hence, the original architecture cannot fit inside the programmable CC engine.

\subsection{Network Quantization}
In an effort to satisfy the low inference time constraints, we began with integer quantization \cite{quantization} and approximated nonlinear activations using lookup tables. Keeping the LSTM element, we then trained policies with smaller architectures and tested their decision time on the device. With these efforts, we reduced the decision latency to approximately $450 \mu$ seconds. Finally, we replaced the LSTM layer with a sliding window over the input history, leading to an MLP with a single hidden layer. In this second attempt, we were able to reduce the decision latency to $17 \mu$ sec. The results are summarized in Table~\ref{table: operations}. Despite this effort, we could not satisfy the desired $2 \mu$sec limit before distilling with tree boosting. 


\begin{figure}
    \centering
    \includegraphics[width=0.7\linewidth]{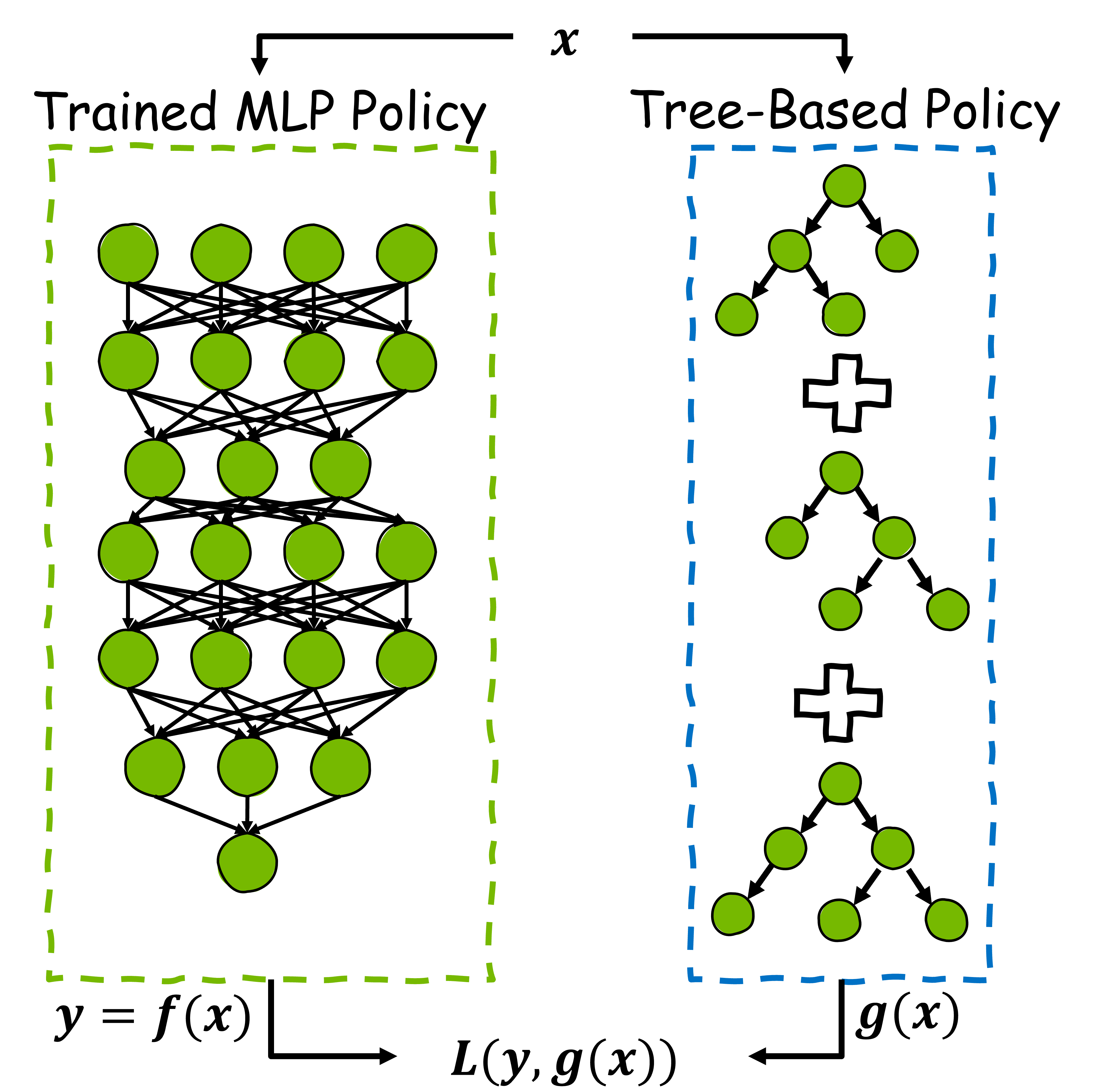}
    \caption{\textbf{Model Distillation}: An illustration of how we train the tree-based student policy $g$ to mimic the fixed NN-based policy $f$ by minimizing $L(y, g(x)) = \sqrt{\frac{1}{N}\sum_{n=1}^N(y_i - g(x_i))^2}$.}
    \label{fig: model distillation}
\end{figure}

\subsection{Boosting Trees}

During our experiments, we were unable to further optimize the NN architecture without harming performance. \cref{table: mlp_vs_tree_deployed} summarizes the failure of the MLP and LSTM architectures due to their long inference time. Hence, to meet the required speed, we opt for binary decision trees.

Learning an optimal RL policy requires continual interaction with an environment. NNs handle this task by gradually updating their parameters through stochastic gradient descent. NN-based policies often require millions or even billions of interactions \cite{badia2020agent57} to achieve an optimal policy. These interactions are conducted with sub-optimal policies during the learning stage, and mostly do not reflect optimal behavior. However, once the policy has converged, learning to imitate it is a supervised learning task. We refer to this process as model distillation \cite{tree_ex_1,tree_ex_2,tree_ex_3,Song_2021_CVPR,tree_ex_rl,DBLP:journals/corr/RusuCGDKPMKH15,tessler2017deep}. \citet{metis2019} highlighted this issue and proposed Metis, a framework for converting NN-based policies to lightweight interpretable controllers based on decision trees. However, in our attempts, a small decision tree fitting within ConnectX-6Dx's memory constraints did not perform well.

Boosting Trees is a suite of ML algorithms used to learn a binary decision tree. They tend to perform extremely well on ML challenges \cite{supervised_learning_comp,tree_vs_ann_2005,boost_vs_dl_2018,algo_comp_classification}, are deterministic, and exhibit robust behavior \cite{trees_robust}. 
We specifically consider Gradient Boosting Trees (GBT) \cite{natekin2013gradient}, an ensemble technique that iteratively adds weak learners (trees) to construct a strong global model. As such, our goal is to distill our NN policy into an equivalent representation using gradient-boosted trees. This process is illustrated in \cref{fig: model distillation}.

Similarly to our definition of the RL policy, our distillation task is to estimate a function $F^*: \mathbb{R}^m \rightarrow \mathbb{R}$, mapping from a set of features to the output action (scalar), a regression problem. To construct this dataset, we collect several trajectories using a convergent policy and record the observed input features and predicted actions. This is done by minimizing an loss function $L(y, F(x))$ in expectation over a training dataset such that
\begin{equation*}
\hat{F} = \argmin_F \E[L(y, F(x))] \,.
\end{equation*}

\begin{table}[t]
    \begin{tabular}{c|cc|cc|cc}
        \textbf{Num. Flows} & \multicolumn{2}{c|}{\textbf{64}} & \multicolumn{2}{c|}{\textbf{512}} & \multicolumn{2}{c}{\textbf{2048}} \\
        \cline{1-1}
        \textbf{Metric} & GP & Latency & GP & Latency & GP & Latency \\ \hline
        \textbf{MLP} & 0.92 & 12.19 & 0.90 & 17.82 & 0.90 & 27.62 \\
        \textbf{Tree (ours)} & 0.92 & 12.03 & 0.90 & 17.98 & 0.90 & 27.35 \\ \hline
    \end{tabular}
    \caption{\textbf{Policy Distillation: Comparing MLP based policy vs distilled policy} on many-to-one scenarios in \textbf{simulation} while varying the number of flows. The simulator emulates a decision latency (inference) of 2 $\mu$sec for both methods. We measure the normalized goodput (GP), and latency (measured in $\mu sec$). These results show that distilling the agent, into gradient-boosted trees, does not degrade the performance.}
    \label{table: mlp_vs_tree_sim}
\end{table}

More specifically, $\hat{F}$ is constructed iteratively as a sequence of estimators such that at each iteration $t$,
$F^t = F^{t-1} + \alpha h^t$, where $\alpha$ is the step-size and $h^t: \mathbb{R}^m \rightarrow \mathbb{R}$ is called the base-predictor. Moreover, $h^t$ is chosen such that
\begin{equation*}
h^t = \argmin_{h \in H} \E[L(y, F^{t-1}(x) + h(x))] \,.
\end{equation*}
The biggest challenge for distillation is to capture the temporal information incorporated by the LSTM layer. Therefore, our first step was to replace the LSTM layer with an MLP before distilling the policy with GBTs.

We chose CatBoost \cite{catboost}, a SOTA GBT implementation in which each $h^t$ is a binary decision tree. Specifically, we restricted the number of boosting iterations and maximal tree depth per tree to satisfy the limits of ConnectX-6Dx. The resulting number of operations does not exceed 150.

In \cref{table: mlp_vs_tree_sim}, we compare the performance differences between the MLP-based teacher model and our tree-based distilled student model. Our results show that using the distillation method, the student is capable of perfectly imitating the performance of the more complex teacher model. Moreover, as shown in \cref{table: operations}, by changing the function class from NN to binary decision trees, we obtained a x500 speed-up, from $450 \mu sec$ down to $0.9 \mu sec$.


\section{Experiments} 
\label{sec: experiements}
Recently, NVIDIA released Programmable Congestion Control (PCC) with its ConnectX-6Dx NIC \cite{pcc}. PCC enables software-based CC to run directly in the networking layer, on dedicated RISC processors, within the NIC. Thanks to this mechanism, we can run our custom CC algorithm after properly representing it in a simple if-else logical structure, which a tree-based policy indeed satisfy. We use PCC to deploy our tree-based policy on a live cluster.

\subsection{Live Cluster Setup} \label{sec:setup}

Our cluster setup involves ConnectX-6Dx NICs connected through a Spectrum-2 switch over a lossy network with a link rate of 100 Gbps. We focus on RoCEv2, an RDMA protocol that runs over Ethernet. For steady-state experiments, we performed inter-rack-traffic tests on a cluster consisting of a two-level Fat-Tree \cite{fat_tree} topology with two spines connected via four 100 Gbps links to four Top of the racks (ToRs) each with 16 nodes. For the last set of reaction experiments (``long-short''), we performed single-rack traffic tests on a single switch cluster with seven hosts.
We generated traffic by continuously posting 64KB RDMA write requests to the receiver.

We compare RL-CC to the official DCQCN implementation, and our best-effort implementation of Swift\footnote{Lacking an official implementation of Swift, we compare to our own best-effort implementation of Swift, deployed on a ConnectX-6Dx device using PCC.}. We trained RL-CC in a single-switch OMNeT++ simulation on various many-to-one and all-to-all scenarios, with the parameters set to $\text{\bf target}=0.064, \beta=1.5$. We then distilled the trained policy, as presented in the previous section, with up to 10 boosting trees of a maximal depth of four.

We compared the CC algorithms' ability to maintain a steady state and react to network changes. For steady-state performance, we evaluated many-to-one, all-to-all, and OSU all-to-all. Here, the goal of the CC is to maximize goodput, while minimizing the latency and packet loss. In addition, we evaluated the reaction time in a long-short test.

\begin{figure}
    \centering 
\caption*{\textbf{many-to-one}}
 \includegraphics[width=\linewidth]{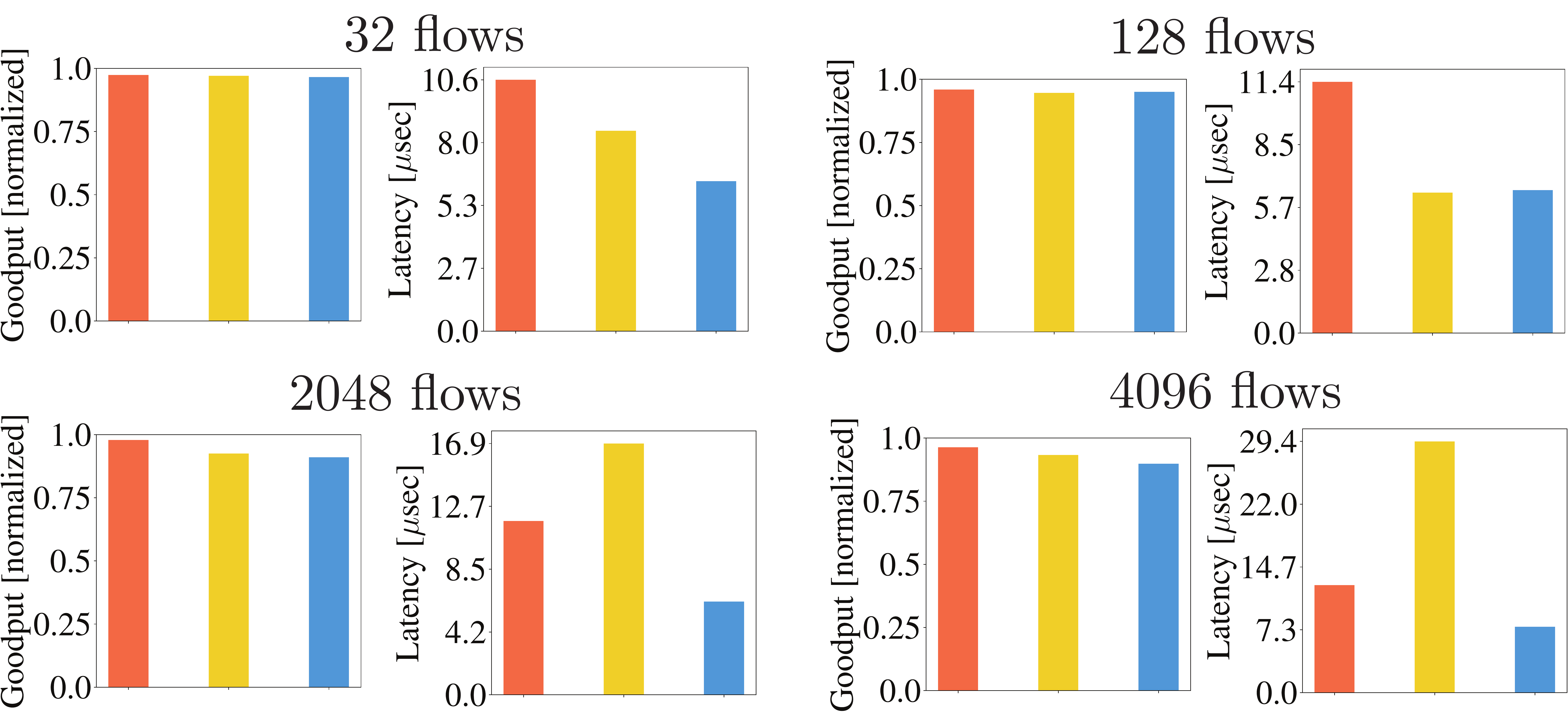}
\caption*{\colorbox{BurntOrange}{\rule[5pt]{5pt}{0pt}} DCQCN \, \colorbox{Yellow}{\rule[5pt]{5pt}{0pt}} Swift \, 
\colorbox{CornflowerBlue}{\rule[5pt]{5pt}{0pt}} RL-CC} 
\caption{Two level fat tree 64-host cluster tests. Test duration is 60 sec for both test. Goodput is normalized to line-rate (higher is better). Latency is measured in $\mu sec$ (lower is better).}
\label{fig:hercules cluster m2o}
\end{figure}

\begin{figure}
    \centering 
\caption*{\textbf{all-to-all}}
 \includegraphics[width=\linewidth]{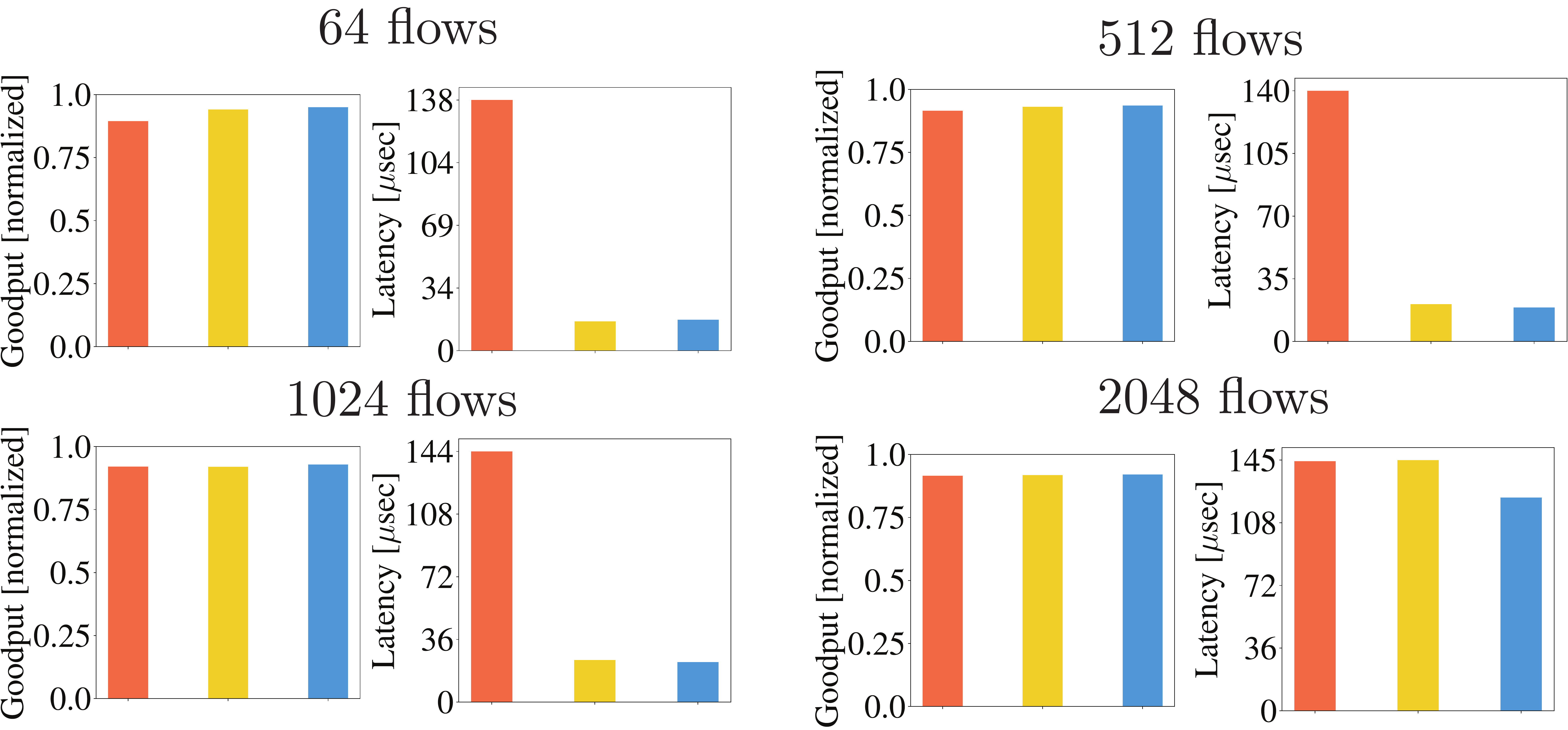}
\caption*{\colorbox{BurntOrange}{\rule[5pt]{5pt}{0pt}} DCQCN \, \colorbox{Yellow}{\rule[5pt]{5pt}{0pt}} Swift \, 
\colorbox{CornflowerBlue}{\rule[5pt]{5pt}{0pt}} RL-CC} 
\caption{Two-level Fat-Tree 64-host cluster tests. Test duration is 60 sec for both test. Goodput is normalized to line-rate (higher is better). Latency is measured in $\mu sec$ (lower is better).}
\label{fig:hercules cluster a2a}
\end{figure}

\subsection{Many-to-One}

Many-to-One evaluates a multiple-sender-single-receiver setup. Multiple hosts, each with multiple active flows, transmit data towards a joint receiver. As all senders share the same destination, they also share the same congestion point. 

We present the results in \cref{fig:hercules cluster m2o}, comparing RL-CC with DCQCN \cite{DCQCN} and Swift \cite{SWIFT}. We present four representative scenarios covering various scales of participating flows. For each scenario, we measure the goodput and the latency. Goodput, as opposed to bandwidth, is the average transmission rate in the network, across all hosts, disregarding re-transmissions due to packet loss. Latency measures the average delay within the network caused by increased congestion.

Here, we observe that while DCQCN attains slightly higher goodput, RL-CC produces similar results, but with a dramatically lower latency. As such, RL-CC is able to reduce congestion across several magnitudes of scale while maximizing network utilization.

\subsection{All-to-All}

All-to-All extends many-to-one to a more chaotic system. Here, multiple hosts run multiple parallel flows. Each host transmits packets to all other hosts. While this is a steady-state test, it is harder to minimize latency as there are multiple congestion points in parallel.

Although DCQCN exhibited extra-ordinary behavior in many-to-one tests, as seen in \cref{fig:hercules cluster a2a}, it underperforms when the system becomes chaotic. Specifically, we observe that RL-CC produces higher goodput and lower latency in all tested scenarios.

\begin{figure}
\centering 
 \includegraphics[width=\linewidth]{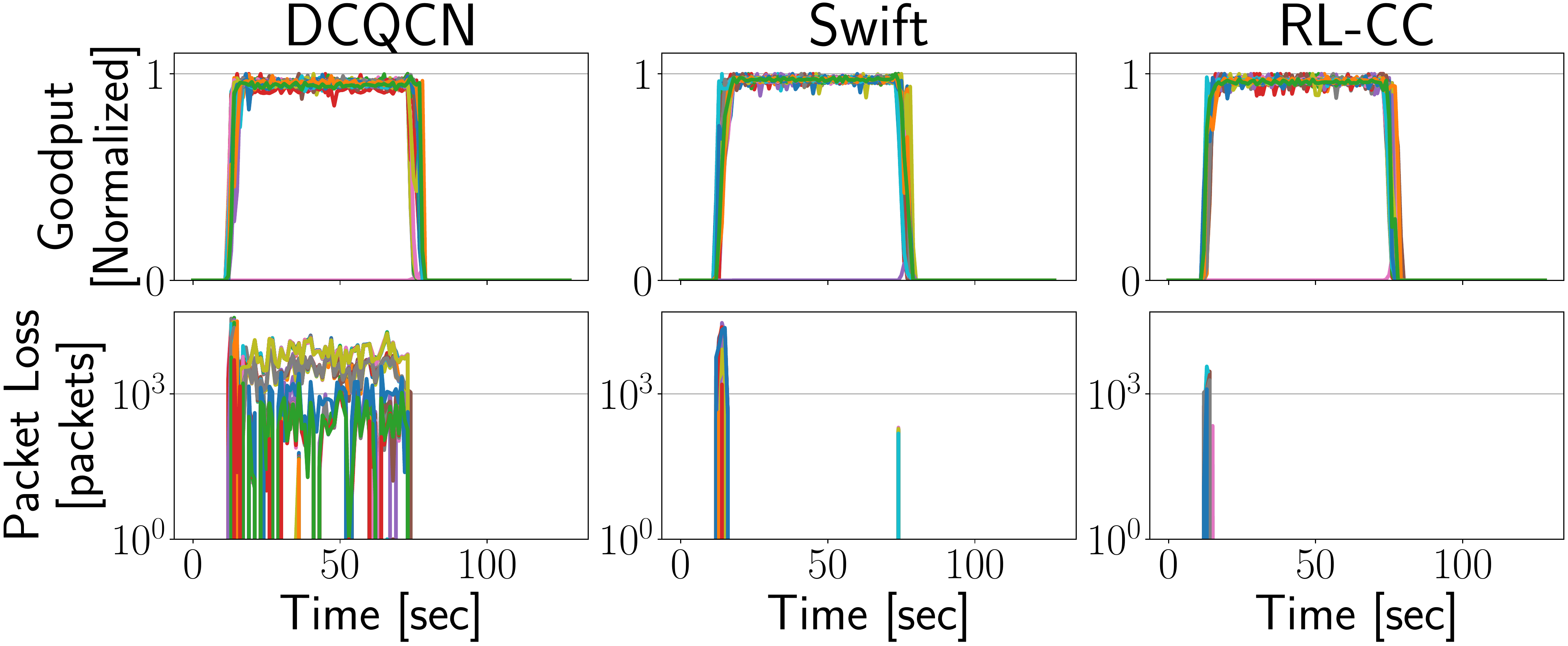}
\caption{Two-level Fat-Tree 64-host cluster -- \textbf{all-to-all} test with 8192 flows. Data was sampled every 1 sec. Different colors denote different hosts.}
\label{fig:hercules cluster 8192_a2a}
\end{figure}

In addition, in \cref{fig:hercules cluster 8192_a2a}, we present the behavior over time. We observe that while RL-CC and Swift successfully control congestion, and only suffer packet loss during the bring-up phase, DCQCN continually fails to control congestion. This is seen by observing the continued packet loss throughout the experiment.

\subsection{OSU All-to-All}

The OSU all-to-all test \cite{PANDA2021101208} measures the latency of the Message Passing Interface (MPI) \cite{mpi40} All-to-All blocking collective across N processes. The test is performed for various message sizes over many iterations. The resulting average latency is the message transmission completion time, which is increased by low bandwidths and packet loss. Therefore, the lower the average latency, the better.  

\begin{figure}
    \centering 
\caption*{\textbf{OSU all-to-all}}
 \includegraphics[width=0.75\linewidth]{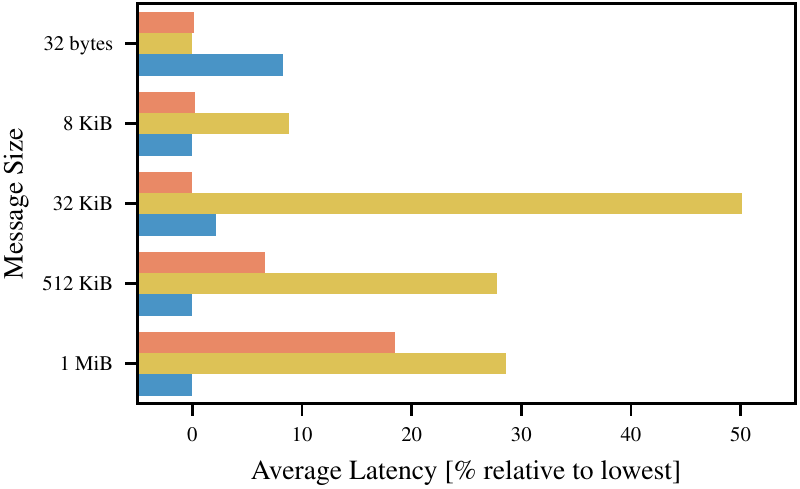}
\caption*{\colorbox{BurntOrange}{\rule[5pt]{5pt}{0pt}} DCQCN \, \colorbox{Yellow}{\rule[5pt]{5pt}{0pt}} Swift \, 
\colorbox{CornflowerBlue}{\rule[5pt]{5pt}{0pt}} RL-CC} 
\caption{OSU all-to-all test on two-level Fat-Tree 64-host cluster (single process per node). The length of the bar represents the average latency relative to the lowest result per message size (lower is better).}
\label{fig:hercules osu a2a}
\end{figure}

As seen in \cref{fig:hercules osu a2a}, with the exception of 32 bytes, RL-CC consistently achieved the lowest latency whereas Swift performed poorly. We attribute this behavior to RL-CC's fast reaction; at larger message sizes, the system quickly enters a stage of congestion, which RL-CC is able to rapidly mitigate. However, when considering tiny packets, the RL-CC reacts too fast, resulting in an unneeded reduction in transmission rate and a longer completion time.

\begin{figure}
    \centering 
\caption*{1 Long flow}
\begin{subfigure}{\linewidth}
  \includegraphics[width=\linewidth]{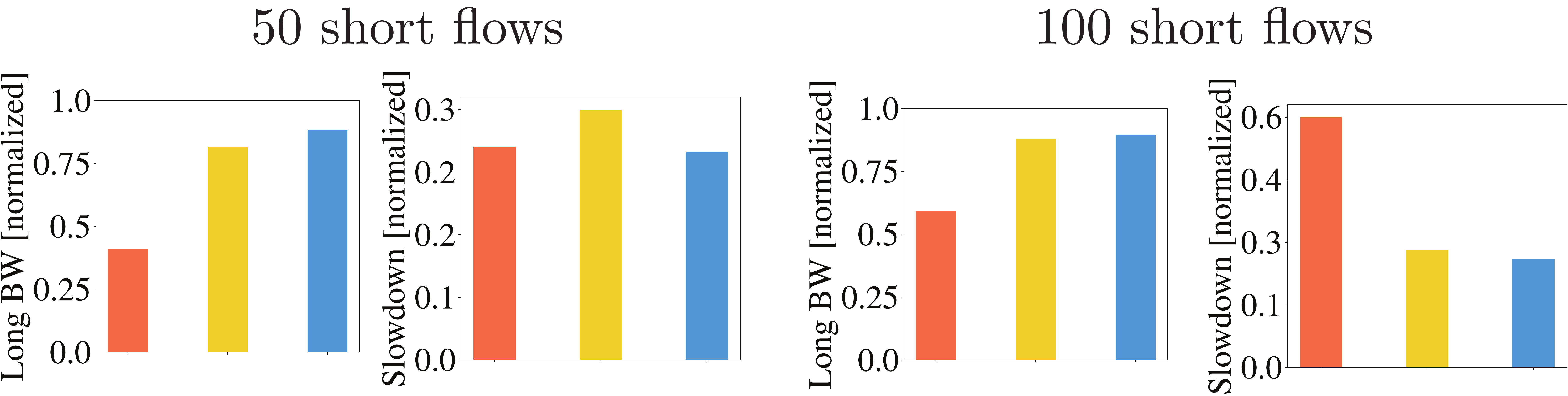}
  \label{subfig: m2o_local_subplots}
\end{subfigure}\hfil 
\caption*{4 Long flows}
\begin{subfigure}{\linewidth}
  \includegraphics[width=\linewidth]{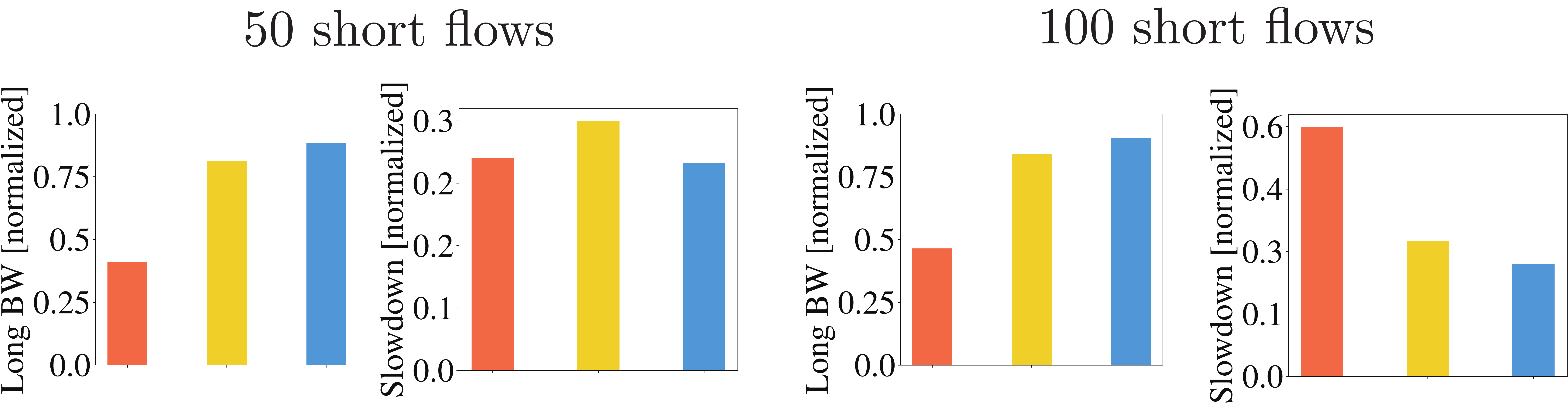}

  \label{subfig: longshort}
\end{subfigure}\hfil 
\caption*{\colorbox{BurntOrange}{\rule[5pt]{5pt}{0pt}} DCQCN \, \colorbox{Yellow}{\rule[5pt]{5pt}{0pt}} Swift \, 
\colorbox{CornflowerBlue}{\rule[5pt]{5pt}{0pt}} RL-CC} 
\caption{\textbf{long-short:} single switch seven-host cluster tests. Test duration is 30 sec for all scenarios. Long BW is normalized to line-rate (higher is better), slowdown is completion time normalized to base RTT (lower is better).}
\label{fig:local cluster long short}
\end{figure}

\begin{table*}[htbp!]
\center
\begin{tabular}{c|cccc|cccc|cc}
\textbf{Scenario}                    & \multicolumn{4}{c|}{\textbf{many-to-one}}                                                  & \multicolumn{4}{c|}{\textbf{all-to-all}}                                                   & \multicolumn{2}{c}{\textbf{long-short}} \\ \hline
\multirow{2}{*}{\textbf{Num. Flows}} & \multirow{2}{*}{32} & \multirow{2}{*}{128} & \multirow{2}{*}{2048} & \multirow{2}{*}{4096} & \multirow{2}{*}{64} & \multirow{2}{*}{512} & \multirow{2}{*}{1024} & \multirow{2}{*}{2048} & \multicolumn{2}{c}{4 long}              \\
                                     &                     &                      &                       &                       &                     &                      &                       &                       & 50 short           & 100 short          \\ \hline
\textbf{DCQCN}                       & \textbf{0.0}        & 886.4                & 28.5                  & 25.7                  & 225.1               & 50.71                & 48.1                  & 26.7                  & \textbf{0.0}       & 57.9               \\
\textbf{Swift}                       & \textbf{0.0}        & \textbf{0.0}         & 10.1                  & 9.1                   & \textbf{0.0}        & \textbf{0.0}         & 0.8                   & 6.5                   & \textbf{0.0}       & \textbf{0.0}       \\
\textbf{RL-CC}                       & \textbf{0.0}        & \textbf{0.0}         & \textbf{0.8}          & \textbf{2.8}          & \textbf{0.0}        & \textbf{0.0}         & \textbf{0.2}          & \textbf{1.2}          & \textbf{0.0}       & \textbf{0.0}       \\ \hline
\end{tabular}
\caption{\textbf{Average amount of packets lost per flow throughout the test} (lower is better). We tested the many-to-one and all-to-all scenarios on a Two-level Fat-Tree 64-hot cluster, whereas the long-short scenario was evaluated on a single switch 7 host cluster. The best results, for each scenario, are highlighted in bold. RL-CC achieves minimal packet loss across all tested scenarios.}
\label{table:packet_loss}
\end{table*}

\subsection{Long-Short}

While the previous tests evaluated the performance during steady state, long-short considers the reaction time. Here, a small number of long flows are continually transmitting data. Then, at a random time during the test, a large number of short flows start transmitting a small amount of data.

We measured both the average long-flow BW and the slowdown. When the short flows begin to transmit, the long flow must reduce its transmission rate and allow the short flows to take part. The reaction time is measured by the slowdown; a higher slowdown means a slower reaction time. On the other hand, once the short flows end their transmission, the long flow should rapidly recover to the full line rate. A faster recovery corresponds to higher long BW.

The results are presented in \cref{fig:local cluster long short}. In all tested scenarios, RL-CC consistently outperforms the baselines, producing better results on both reaction and recovery metrics.

\subsection{Packet Loss}
In the above analysis, we focus on metrics such as goodput and latency. We now inspect a complementary measurement for all tests, packet loss. Packet loss occurs when the CC algorithm is too slow to react. The various flows transmit at a too high rate, filling the network queues, and resulting in packet loss and network performance degradation.

We give the results in \cref{table:packet_loss}. They further emphasize that RL-CC packet losses occur during the bring-up phase in steady-state scenarios. Furthermore, this picture complements the long-short experiment. The fast reaction time of RL-CC is highlighted by its ability to minimize packet loss across all scenarios.

\subsection{Explainable Reinforcement Learning}\label{subsec: explainable rl}

\begin{table*}
    \large
    \begin{center}
    \begin{tabular}{c|cccc}
         & \textbf{many-to-one} & \textbf{all-to-all}  & \textbf{OSU} & \textbf{long-short} \\ \hline
        DCQCN & \color{ForestGreen}{\cmark} & \color{RubineRed}{\xmark}   & \color{ForestGreen}{\cmark} & \color{RubineRed}{\xmark} \\
        Swift & \color{RubineRed}{\xmark} & \color{ForestGreen}{\cmark} & \color{RubineRed}{\xmark} & \color{ForestGreen}{\cmark} \\
        \textbf{RL-CC} & \color{ForestGreen}{\cmark} & \color{ForestGreen}{\cmark} & \color{ForestGreen}{\cmark} & \color{ForestGreen}{\cmark} \\ \hline
    \end{tabular}
    \caption{Comparison of various approaches. A {\cmark} means the method has successfully controlled and prevented congestion in this task, whereas {\xmark} presents a failure. As can be seen, RL-CC is the only CC algorithm, among the compared methods, that succeeds in all tasks.}
    \end{center}
\end{table*}

In previous sections, we analyzed the performance of RL-CC and compared it with standard practices used in production -- DCQCN and Swift.
These were designed by humans with interpretable rules, whereas RL-CC's logic is learned from data. Here, we analyze RL-CC's decision-making process. We believe that this not only provides insight into RL-CC but also can provide insight for future rule-based methods.



\begin{table}[t]
    \centering
    \begin{tabular}{c|c|c|c|c}
         & \multicolumn{1}{c}{} & \multicolumn{3}{c}{\textbf{Current system condition}} \\
        \hline
        & & Under-utilized & On target & Congested \\
        \cline{2-5}
        \multirow{3}{1.3cm}{\textbf{Previous system condition}} & Under-utilized & \cellcolor{red!15} $1.05$ & \cellcolor{blue!15} $0.92$ & \cellcolor{blue!30} $0.89$ \\
        \cline{2-5}
        & On target & \cellcolor{red!23} $1.1$ & $1$ & \cellcolor{blue!30} $0.9$ \\
        \cline{2-5}
        & Congested & \cellcolor{red!30} $1.15$ & \cellcolor{red!19} $1.07$ & \cellcolor{blue!10} $0.94$
    \end{tabular}
    \caption{\textbf{Analyzing the logic behind RL-CC:} We illustrate how RL-CC reacts to changes in the system. Each cell represents a combination of previous and current system conditions. Its value is  RL-CC's action -- the multiplicative transmission rate increase/decrease.}
    \label{tab: interpreting rl-cc}
\end{table}

\begin{figure}[t]
    \centering
    \includegraphics[width=\linewidth]{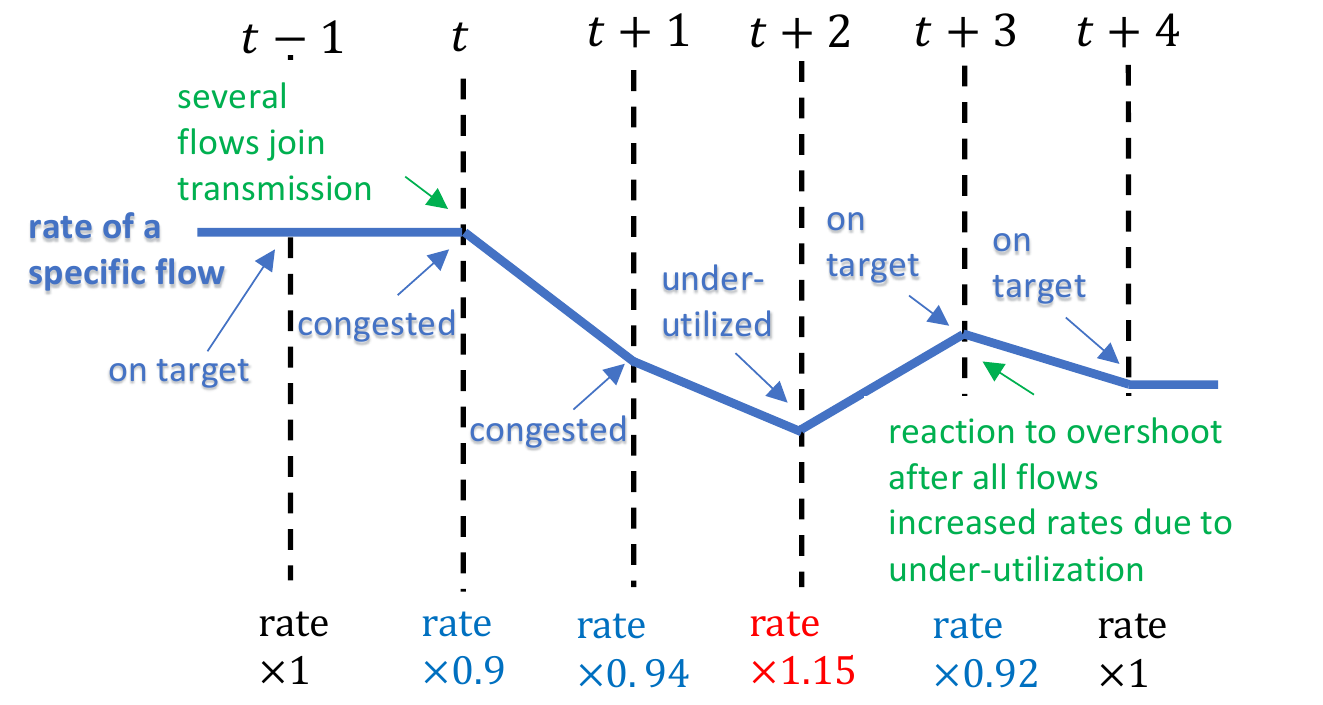}
    \caption{\textbf{Explaining RL:} A hypothetical scenario demonstrating the behavior depicted in \cref{tab: interpreting rl-cc}. The following description is our interpretation of the numerical outputs we observe in the policy: When congestion starts due to new transmission, RL-CC decreases the rate rapidly for fast reaction, but afterward dampens the rate change due the the inertia-like property of the system. Then, the line becomes under-utilized so RL-CC increases the rate and the target is achieved. But, despite hitting target, as opposed to other CC algorithms, it then reduces the rate slightly; it predicts that other flows will behave similarly so it applies a `stabilizing' action. }
    \label{fig: explanable scenario}
\end{figure}

To study the logic behind RL-CC, we input nine combinations of artificial values of past and current states and measured the respective outputs. \cref{tab: interpreting rl-cc} summarizes the results. As RL-CC is a history-dependent algorithm, the rows represent the previous system state, and the columns the current system state. We begin with ``first-order'' reactions: (a) when the system is under-utilized, RL-CC raises the transmission rate (left column); (b) when the system is at the target delay, RL-CC barely changes the rate (middle column); and (c) if the system is in a state of congestion, RL-CC reduces the transmission rate (right column).

In addition, we observe that RL-CC also learned non-trivial second-order reactions.
For instance, when the system initially becomes congested (or, alternatively, under-utilized), RL-CC reduces (increases) the transmission rate more harshly than when the system remains congested (under-utilized) for multiple steps. Similarly, when the system transitions from under-utilized to being on target, RL-CC is proactive and decreases the transmission rate. This suggests that RL-CC learned the inertia-like behavior of the system and anticipates the reaction of its own flow together with that of other agents operating in parallel. This behavior is exactly the opposite of DCQCN's increment and decrement scheme, which becomes more aggressive in a similar situation. When DCQCN receives a congestion notification, it multiplicatively reduces the rate by the variable $\alpha$ which increases as the indications continue to come. Increments are triggered by a timer and byte counter and also become more aggressive as consecutive triggers occur.

In \cref{fig: explanable scenario}, we illustrate the logic behind \cref{tab: interpreting rl-cc} with a hypothetical scenario of how the system stabilizes after reacting to congestion.

\section{Conclusion and Future Directions}

Effective CC is crucial for high network performance in modern datacenters. The benefit of ML methods lies in their ability to extract meaningful patterns from complex data, often making them better than humans in such tasks. To the best of our knowledge, we presented here for the first time in the literature an ML-based CC method that can successfully run in real time within operational datacenters.

On the path to deployment, we began with an extensive analysis of RL-CC. We performed a thorough study of the various trade-offs in the reward design and showed that, in contrast to previous methods, RL-CC is capable of precisely tracking the optimal theoretical inflation curves.

The agent, initially represented using a NN, required $450 \mu sec$ to perform inference. Due to the rate of change within the datacenter, this latency was too high and resulted in the inability to control congestion. To overcome this challenge, we showed that by changing the architecture (from an LSTM to a sliding-window MLP), the learned policy can be distilled into decision trees. This resulted in a reduction of inference time to $0.9 \mu sec$, an improvement of x500.

We then deploy RL-CC on a real cluster consisting of 64 hosts. In these tests, the policy ran in real-time directly on ConnectX-6Dx NICs. RL-CC demonstrated high goodput and fairness while retaining low packet latency and minimal packet loss. Moreover, we showed the ability of RL-CC to generalize, out of the box, to new and unseen scenarios.

Finally, we provided insights into RL-CC's decision-making process. We inspected the output sensitivity to combinations of prior and present states. Surprisingly, RL-CC not only learned expected reactive behaviors, but also learned to anticipate via second-order predictions. This analysis sheds light on the feasibility of a data-driven MIMD approach, challenging the previous belief that AIMD is required to converge to a stable and fair solution.

Our tree-based RL-CC is an initial step towards real-world lightweight AI CC. AI methods generally perform better when trained on larger and richer data. In future work, we aim to study additional network signals that may enable a better prediction of the network state. 

\bibliographystyle{IEEEtranN}
\bibliography{bibliography}

\begin{thebibliography}{49}
\providecommand{\natexlab}[1]{#1}
\providecommand{\url}[1]{#1}
\csname url@samestyle\endcsname
\providecommand{\newblock}{\relax}
\providecommand{\bibinfo}[2]{#2}
\providecommand{\BIBentrySTDinterwordspacing}{\spaceskip=0pt\relax}
\providecommand{\BIBentryALTinterwordstretchfactor}{4}
\providecommand{\BIBentryALTinterwordspacing}{\spaceskip=\fontdimen2\font plus
\BIBentryALTinterwordstretchfactor\fontdimen3\font minus
  \fontdimen4\font\relax}
\providecommand{\BIBforeignlanguage}[2]{{%
\expandafter\ifx\csname l@#1\endcsname\relax
\typeout{** WARNING: IEEEtranN.bst: No hyphenation pattern has been}%
\typeout{** loaded for the language `#1'. Using the pattern for}%
\typeout{** the default language instead.}%
\else
\language=\csname l@#1\endcsname
\fi
#2}}
\providecommand{\BIBdecl}{\relax}
\BIBdecl

\bibitem[Tessler et~al.(2022)Tessler, Shpigelman, Dalal, Mandelbaum,
  Haritan~Kazakov, Fuhrer, Chechik, and Mannor]{rl-cc}
C.~Tessler, Y.~Shpigelman, G.~Dalal, A.~Mandelbaum, D.~Haritan~Kazakov,
  B.~Fuhrer, G.~Chechik, and S.~Mannor, ``Reinforcement learning for datacenter
  congestion control,'' \emph{Proceedings of the AAAI Conference on Artificial
  Intelligence}, vol.~36, no.~11, pp. 12\,615--12\,621, Jun. 2022.

\bibitem[Burstein(2021)]{burstein2021nvidia}
I.~Burstein, ``Nvidia data center processing unit (dpu) architecture,'' in
  \emph{2021 IEEE Hot Chips 33 Symposium (HCS)}.\hskip 1em plus 0.5em minus
  0.4em\relax IEEE, 2021, pp. 1--20.

\bibitem[Beck and Kagan(2011)]{beck2011performance}
M.~Beck and M.~Kagan, ``Performance evaluation of the rdma over ethernet (roce)
  standard in enterprise data centers infrastructure,'' in \emph{Proceedings of
  the 3rd Workshop on Data Center-Converged and Virtual Ethernet Switching},
  2011, pp. 9--15.

\bibitem[Guo et~al.(2016)Guo, Wu, Deng, Soni, Ye, Padhye, and
  Lipshteyn]{rdma_widespread_use}
C.~Guo, H.~Wu, Z.~Deng, G.~Soni, J.~Ye, J.~Padhye, and M.~Lipshteyn, ``Rdma
  over commodity ethernet at scale,'' in \emph{Proceedings of the 2016 ACM
  SIGCOMM Conference}, ser. SIGCOMM '16.\hskip 1em plus 0.5em minus 0.4em\relax
  New York, NY, USA: Association for Computing Machinery, 2016, p. 202–215.

\bibitem[Zhu et~al.(2015)Zhu, Eran, Firestone, Guo, Lipshteyn, Liron, Padhye,
  Raindel, Yahia, and Zhang]{DCQCN}
Y.~Zhu, H.~Eran, D.~Firestone, C.~Guo, M.~Lipshteyn, Y.~Liron, J.~Padhye,
  S.~Raindel, M.~H. Yahia, and M.~Zhang, ``Congestion control for large-scale
  rdma deployments,'' in \emph{Proceedings of the 2015 ACM Conference on
  Special Interest Group on Data Communication}, ser. SIGCOMM '15.\hskip 1em
  plus 0.5em minus 0.4em\relax New York, NY, USA: Association for Computing
  Machinery, 2015, p. 523–536.

\bibitem[Kumar et~al.(2020)Kumar, Dukkipati, Jang, Wassel, Wu, Montazeri, Wang,
  Springborn, Alfeld, Ryan, Wetherall, and Vahdat]{SWIFT}
G.~Kumar, N.~Dukkipati, K.~Jang, H.~M.~G. Wassel, X.~Wu, B.~Montazeri, Y.~Wang,
  K.~Springborn, C.~Alfeld, M.~Ryan, D.~Wetherall, and A.~Vahdat, ``Swift:
  Delay is simple and effective for congestion control in the datacenter,'' in
  \emph{Proceedings of the Annual Conference of the ACM Special Interest Group
  on Data Communication on the Applications, Technologies, Architectures, and
  Protocols for Computer Communication}, ser. SIGCOMM '20.\hskip 1em plus 0.5em
  minus 0.4em\relax New York, NY, USA: Association for Computing Machinery,
  2020, p. 514–528.

\bibitem[Shpigelman et~al.(U.S. Patent 0152474 A1 , May, 2021)Shpigelman,
  Burstein, Bloch, Zuck, and Moyal]{pcc}
Y.~Shpigelman, I.~Burstein, N.~Bloch, R.~Zuck, and R.~Moyal, ``Programmable
  congestion control communication scheme,'' U.S. Patent 0152474 A1 , May,
  2021.

\bibitem[Shpiner et~al.(2017)Shpiner, Zahavi, Dahley, Barnea, Damsker, Yekelis,
  Zus, Kuta, and Baram]{rocerocks}
A.~Shpiner, E.~Zahavi, O.~Dahley, A.~Barnea, R.~Damsker, G.~Yekelis, M.~Zus,
  E.~Kuta, and D.~Baram, ``Roce rocks without pfc: Detailed evaluation,'' in
  \emph{Proceedings of the Workshop on Kernel-Bypass Networks}, ser. KBNets
  '17.\hskip 1em plus 0.5em minus 0.4em\relax New York, NY, USA: Association
  for Computing Machinery, 2017, p. 25–30.

\bibitem[Hu et~al.(2016)Hu, Zhu, Cheng, Guo, Tan, Padhye, and
  Chen]{pfc_deadlock}
S.~Hu, Y.~Zhu, P.~Cheng, C.~Guo, K.~Tan, J.~Padhye, and K.~Chen, ``Deadlocks in
  datacenter networks: Why do they form, and how to avoid them,'' in
  \emph{Proceedings of the 15th ACM Workshop on Hot Topics in Networks}, ser.
  HotNets '16.\hskip 1em plus 0.5em minus 0.4em\relax New York, NY, USA:
  Association for Computing Machinery, 2016, p. 92–98.

\bibitem[Bui et~al.(2021)Bui, Van~Chien, Lagunas, Grotz, Chatzinotas, and
  Ottersten]{RoCC}
V.-P. Bui, T.~Van~Chien, E.~Lagunas, J.~Grotz, S.~Chatzinotas, and
  B.~Ottersten, ``Robust congestion control for demand-based optimization in
  precoded multi-beam high throughput satellite communications,'' 2021.

\bibitem[Mittal et~al.(2015)Mittal, Lam, Dukkipati, Blem, Wassel, Ghobadi,
  Vahdat, Wang, Wetherall, and Zats]{TIMELY}
R.~Mittal, V.~T. Lam, N.~Dukkipati, E.~Blem, H.~Wassel, M.~Ghobadi, A.~Vahdat,
  Y.~Wang, D.~Wetherall, and D.~Zats, ``Timely: Rtt-based congestion control
  for the datacenter,'' \emph{SIGCOMM Comput. Commun. Rev.}, vol.~45, no.~4, p.
  537–550, aug 2015.

\bibitem[Zhang et~al.(2021)Zhang, Liu, Meng, and Ren]{cong_in_lossless}
Y.~Zhang, Y.~Liu, Q.~Meng, and F.~Ren, ``Congestion detection in lossless
  networks,'' in \emph{Proceedings of the 2021 ACM SIGCOMM 2021 Conference},
  ser. SIGCOMM '21.\hskip 1em plus 0.5em minus 0.4em\relax New York, NY, USA:
  Association for Computing Machinery, 2021, p. 370–383.

\bibitem[Yuan et~al.(2021)Yuan, Zhou, Dong, and Huang]{breaking_one_rtt}
G.~Yuan, R.~Zhou, D.~Dong, and S.~Huang, ``Breaking one-rtt barrier:
  Ultra-precise and efficient congestion control in datacenter networks,'' in
  \emph{2021 International Conference on Computer Communications and Networks
  (ICCCN)}, 2021, pp. 1--9.

\bibitem[Zhou et~al.(2021)Zhou, Dong, Huang, and Bai]{fasttune}
R.~Zhou, D.~Dong, S.~Huang, and Y.~Bai, ``Fasttune: Timely and precise
  congestion control in data center network,'' in \emph{2021 IEEE Intl Conf on
  Parallel and Distributed Processing with Applications, Big Data and Cloud
  Computing, Sustainable Computing and Communications, Social Computing and
  Networking (ISPA/BDCloud/SocialCom/SustainCom)}, 2021, pp. 238--245.

\bibitem[Floyd et~al.(2001)Floyd, Ramakrishnan, and Black]{ECN}
S.~Floyd, D.~K.~K. Ramakrishnan, and D.~L. Black, ``{The Addition of Explicit
  Congestion Notification (ECN) to IP},'' RFC 3168, Sep. 2001.

\bibitem[Li et~al.(2019)Li, Miao, Liu, Zhuang, Feng, Tang, Cao, Zhang, Kelly,
  Alizadeh, and Yu]{HPCC}
Y.~Li, R.~Miao, H.~H. Liu, Y.~Zhuang, F.~Feng, L.~Tang, Z.~Cao, M.~Zhang,
  F.~Kelly, M.~Alizadeh, and M.~Yu, ``Hpcc: High precision congestion
  control,'' in \emph{Proceedings of the ACM Special Interest Group on Data
  Communication}, ser. SIGCOMM '19.\hskip 1em plus 0.5em minus 0.4em\relax New
  York, NY, USA: Association for Computing Machinery, 2019, p. 44–58.

\bibitem[Chiu and Jain(1989)]{aimd_convergence}
D.-M. Chiu and R.~Jain, ``Analysis of the increase and decrease algorithms for
  congestion avoidance in computer networks,'' \emph{Computer Networks and ISDN
  Systems}, vol.~17, no.~1, pp. 1--14, 1989.

\bibitem[Floyd(2003)]{highspeed_aimd_fail}
S.~Floyd, ``Rfc3649: Highspeed tcp for large congestion windows,'' 2003.

\bibitem[Jiang et~al.(2020)Jiang, Li, Jiang, Shen, Sinnott, Tian, and
  Xu]{ml_in_cc}
H.~Jiang, Q.~Li, Y.~Jiang, G.~Shen, R.~O. Sinnott, C.~Tian, and M.~Xu, ``When
  machine learning meets congestion control: {A} survey and comparison,''
  \emph{CoRR}, vol. abs/2010.11397, 2020.

\bibitem[Dong et~al.(2018)Dong, Meng, Zarchy, Arslan, Gilad, Godfrey, and
  Schapira]{vivas}
M.~Dong, T.~Meng, D.~Zarchy, E.~Arslan, Y.~Gilad, B.~Godfrey, and M.~Schapira,
  ``{PCC} vivace: {Online-Learning} congestion control,'' in \emph{15th USENIX
  Symposium on Networked Systems Design and Implementation (NSDI 18)}.\hskip
  1em plus 0.5em minus 0.4em\relax Renton, WA: USENIX Association, Apr. 2018,
  pp. 343--356.

\bibitem[Jay et~al.(2019)Jay, Rotman, Godfrey, Schapira, and
  Tamar]{jay2019deep}
N.~Jay, N.~Rotman, B.~Godfrey, M.~Schapira, and A.~Tamar, ``A deep
  reinforcement learning perspective on internet congestion control,'' in
  \emph{International Conference on Machine Learning}.\hskip 1em plus 0.5em
  minus 0.4em\relax PMLR, 2019, pp. 3050--3059.

\bibitem[Jin et~al.(2018)Jin, Li, Tuo, Wang, and Li]{sdn-rl}
R.~Jin, J.~Li, X.~Tuo, W.~Wang, and X.~Li, ``A congestion control method of sdn
  data center based on reinforcement learning,'' \emph{International Journal of
  Communication Systems}, vol.~31, no.~17, p. e3802, 2018, e3802
  IJCS-18-0005.R1.

\bibitem[Watkins and Dayan(1992)]{qlearning}
C.~J. Watkins and P.~Dayan, ``Technical note: Q-learning,'' \emph{Machine
  Learning}, vol.~8, no.~3, pp. 279--292, May 1992.

\bibitem[Rummery and Niranjan(1994)]{sarsa}
G.~Rummery and M.~Niranjan, ``On-line q-learning using connectionist systems,''
  \emph{Technical Report CUED/F-INFENG/TR 166}, 11 1994.

\bibitem[Lan et~al.(2019)Lan, Tan, Lv, Jin, and Yang]{ndn-rl}
D.~Lan, X.~Tan, J.~Lv, Y.~Jin, and J.~Yang, ``A deep reinforcement learning
  based congestion control mechanism for ndn,'' in \emph{ICC 2019 - 2019 IEEE
  International Conference on Communications (ICC)}, 2019, pp. 1--7.

\bibitem[Mai et~al.(2019)Mai, Yao, Jing, Xu, Wang, and Ji]{rl-satellites}
T.~Mai, H.~Yao, Y.~Jing, X.~Xu, X.~Wang, and Z.~Ji, ``Self-learning congestion
  control of mptcp in satellites communications,'' in \emph{2019 15th
  International Wireless Communications \& Mobile Computing Conference
  (IWCMC)}, 2019, pp. 775--780.

\bibitem[Lillicrap et~al.(2015)Lillicrap, Hunt, Pritzel, Heess, Erez, Tassa,
  Silver, and Wierstra]{ddpg}
T.~P. Lillicrap, J.~J. Hunt, A.~Pritzel, N.~Heess, T.~Erez, Y.~Tassa,
  D.~Silver, and D.~Wierstra, ``Continuous control with deep reinforcement
  learning,'' 2015.

\bibitem[Hochreiter and Schmidhuber(1997)]{LSTM}
S.~Hochreiter and J.~Schmidhuber, ``Long short-term memory,'' \emph{Neural
  Comput.}, vol.~9, no.~8, p. 1735–1780, nov 1997.

\bibitem[Varga(2002)]{varga2002OMNeT++}
A.~Varga, ``Omnet++ http://www. omnetpp. org,'' \emph{IEEE Network
  Interactive}, vol.~16, no.~4, 2002.

\bibitem[Wu et~al.(2020)Wu, Judd, Zhang, Isaev, and Micikevicius]{quantization}
H.~Wu, P.~Judd, X.~Zhang, M.~Isaev, and P.~Micikevicius, ``Integer quantization
  for deep learning inference: Principles and empirical evaluation,'' 2020.

\bibitem[Badia et~al.(2020)Badia, Piot, Kapturowski, Sprechmann, Vitvitskyi,
  Guo, and Blundell]{badia2020agent57}
A.~P. Badia, B.~Piot, S.~Kapturowski, P.~Sprechmann, A.~Vitvitskyi, Z.~D. Guo,
  and C.~Blundell, ``Agent57: Outperforming the atari human benchmark,'' in
  \emph{International Conference on Machine Learning}.\hskip 1em plus 0.5em
  minus 0.4em\relax PMLR, 2020, pp. 507--517.

\bibitem[Che et~al.(2016)Che, Purushotham, Khemani, and Liu]{tree_ex_1}
Z.~Che, S.~Purushotham, R.~Khemani, and Y.~Liu, ``Interpretable deep models for
  icu outcome prediction,'' in \emph{AMIA annual symposium proceedings}, vol.
  2016.\hskip 1em plus 0.5em minus 0.4em\relax American Medical Informatics
  Association, 2016, p. 371.

\bibitem[Liu et~al.(2018)Liu, Wang, and Matwin]{tree_ex_2}
X.~Liu, X.~Wang, and S.~Matwin, ``Improving the interpretability of deep neural
  networks with knowledge distillation,'' in \emph{2018 IEEE International
  Conference on Data Mining Workshops (ICDMW)}.\hskip 1em plus 0.5em minus
  0.4em\relax IEEE, 2018, pp. 905--912.

\bibitem[Li et~al.(2020)Li, Li, Xiang, Xia, Dong, and Cai]{tree_ex_3}
J.~Li, Y.~Li, X.~Xiang, S.-T. Xia, S.~Dong, and Y.~Cai, ``Tnt: An interpretable
  tree-network-tree learning framework using knowledge distillation,''
  \emph{Entropy}, vol.~22, no.~11, 2020.

\bibitem[Song et~al.(2021)Song, Zhang, Wang, Xue, Chen, Sun, Tao, and
  Song]{Song_2021_CVPR}
J.~Song, H.~Zhang, X.~Wang, M.~Xue, Y.~Chen, L.~Sun, D.~Tao, and M.~Song,
  ``Tree-like decision distillation,'' in \emph{Proceedings of the IEEE/CVF
  Conference on Computer Vision and Pattern Recognition (CVPR)}, June 2021, pp.
  13\,488--13\,497.

\bibitem[Biggs et~al.(2021)Biggs, Sun, and Ettl]{tree_ex_rl}
M.~Biggs, W.~Sun, and M.~Ettl, ``Model distillation for revenue optimization:
  Interpretable personalized pricing,'' in \emph{International Conference on
  Machine Learning}.\hskip 1em plus 0.5em minus 0.4em\relax PMLR, 2021, pp.
  946--956.

\bibitem[Rusu et~al.(2016)Rusu, Colmenarejo, Çaglar Gülçehre, Desjardins,
  Kirkpatrick, Pascanu, Mnih, Kavukcuoglu, and
  Hadsell]{DBLP:journals/corr/RusuCGDKPMKH15}
\BIBentryALTinterwordspacing
A.~A. Rusu, S.~G. Colmenarejo, Çaglar Gülçehre, G.~Desjardins,
  J.~Kirkpatrick, R.~Pascanu, V.~Mnih, K.~Kavukcuoglu, and R.~Hadsell, ``Policy
  distillation,'' in \emph{ICLR (Poster)}, 2016. [Online]. Available:
  \url{http://arxiv.org/abs/1511.06295}
\BIBentrySTDinterwordspacing

\bibitem[Tessler et~al.(2017)Tessler, Givony, Zahavy, Mankowitz, and
  Mannor]{tessler2017deep}
C.~Tessler, S.~Givony, T.~Zahavy, D.~Mankowitz, and S.~Mannor, ``A deep
  hierarchical approach to lifelong learning in minecraft,'' in
  \emph{Proceedings of the AAAI Conference on Artificial Intelligence},
  vol.~31, no.~1, 2017.

\bibitem[Meng et~al.(2019)Meng, Wang, Xu, Mao, Bai, and Hu]{metis2019}
\BIBentryALTinterwordspacing
Z.~Meng, M.~Wang, M.~Xu, H.~Mao, J.~Bai, and H.~Hu, ``Explaining deep
  learning-based networked systems,'' \emph{CoRR}, vol. abs/1910.03835, 2019.
  [Online]. Available: \url{http://arxiv.org/abs/1910.03835}
\BIBentrySTDinterwordspacing

\bibitem[Caruana and Niculescu-Mizil(2006)]{supervised_learning_comp}
R.~Caruana and A.~Niculescu-Mizil, ``An empirical comparison of supervised
  learning algorithms,'' in \emph{Proceedings of the 23rd International
  Conference on Machine Learning}, ser. ICML '06.\hskip 1em plus 0.5em minus
  0.4em\relax New York, NY, USA: Association for Computing Machinery, 2006, p.
  161–168.

\bibitem[Roe et~al.(2005)Roe, Yang, Zhu, Liu, Stancu, and
  McGregor]{tree_vs_ann_2005}
B.~P. Roe, H.-J. Yang, J.~Zhu, Y.~Liu, I.~Stancu, and G.~McGregor, ``Boosted
  decision trees as an alternative to artificial neural networks for particle
  identification,'' \emph{Nuclear Instruments and Methods in Physics Research
  Section A: Accelerators, Spectrometers, Detectors and Associated Equipment},
  vol. 543, no. 2–3, p. 577–584, May 2005.

\bibitem[Anghel et~al.(2018)Anghel, Cioara, Moldovan, Salomie, and
  Tomus]{boost_vs_dl_2018}
I.~Anghel, T.~Cioara, D.~Moldovan, I.~Salomie, and M.~M. Tomus, ``Prediction of
  manufacturing processes errors: Gradient boosted trees versus deep neural
  networks,'' in \emph{2018 IEEE 16th International Conference on Embedded and
  Ubiquitous Computing (EUC)}, 2018, pp. 29--36.

\bibitem[Zhang et~al.(2017)Zhang, Liu, Zhang, and
  Almpanidis]{algo_comp_classification}
C.~Zhang, C.~Liu, X.~Zhang, and G.~Almpanidis, ``An up-to-date comparison of
  state-of-the-art classification algorithms,'' \emph{Expert Syst. Appl.},
  vol.~82, no.~C, p. 128–150, oct 2017.

\bibitem[Einziger et~al.(2019)Einziger, Goldstein, Sa'ar, and
  Segall]{trees_robust}
G.~Einziger, M.~Goldstein, Y.~Sa'ar, and I.~Segall, ``Verifying robustness of
  gradient boosted models,'' in \emph{Proceedings of the Thirty-Third AAAI
  Conference on Artificial Intelligence and Thirty-First Innovative
  Applications of Artificial Intelligence Conference and Ninth AAAI Symposium
  on Educational Advances in Artificial Intelligence}, ser.
  AAAI'19/IAAI'19/EAAI'19.\hskip 1em plus 0.5em minus 0.4em\relax AAAI Press,
  2019.

\bibitem[Natekin and Knoll(2013)]{natekin2013gradient}
A.~Natekin and A.~Knoll, ``Gradient boosting machines, a tutorial,''
  \emph{Frontiers in neurorobotics}, vol.~7, p.~21, 2013.

\bibitem[Prokhorenkova et~al.(2018)Prokhorenkova, Gusev, Vorobev, Dorogush, and
  Gulin]{catboost}
L.~Prokhorenkova, G.~Gusev, A.~Vorobev, A.~V. Dorogush, and A.~Gulin,
  ``Catboost: unbiased boosting with categorical features,'' in \emph{Advances
  in Neural Information Processing Systems}, S.~Bengio, H.~Wallach,
  H.~Larochelle, K.~Grauman, N.~Cesa-Bianchi, and R.~Garnett, Eds.,
  vol.~31.\hskip 1em plus 0.5em minus 0.4em\relax Curran Associates, Inc.,
  2018.

\bibitem[Al-Fares et~al.(2008)Al-Fares, Loukissas, and Vahdat]{fat_tree}
M.~Al-Fares, A.~Loukissas, and A.~Vahdat, ``A scalable, commodity data center
  network architecture,'' in \emph{Proceedings of the ACM SIGCOMM 2008
  Conference on Data Communication}, ser. SIGCOMM '08.\hskip 1em plus 0.5em
  minus 0.4em\relax New York, NY, USA: Association for Computing Machinery,
  2008, p. 63–74.

\bibitem[Panda et~al.(2021)Panda, Subramoni, Chu, and
  Bayatpour]{PANDA2021101208}
\BIBentryALTinterwordspacing
D.~K. Panda, H.~Subramoni, C.-H. Chu, and M.~Bayatpour, ``The mvapich project:
  Transforming research into high-performance mpi library for hpc community,''
  \emph{Journal of Computational Science}, vol.~52, p. 101208, 2021, case
  Studies in Translational Computer Science. [Online]. Available:
  \url{https://www.sciencedirect.com/science/article/pii/S1877750320305093}
\BIBentrySTDinterwordspacing

\bibitem[{Message Passing Interface Forum}(2021)]{mpi40}
\BIBentryALTinterwordspacing
{Message Passing Interface Forum}, \emph{{MPI}: A Message-Passing Interface
  Standard Version 4.0}, Jun. 2021. [Online]. Available:
  \url{https://www.mpi-forum.org/docs/mpi-4.0/mpi40-report.pdf}
\BIBentrySTDinterwordspacing

\end{thebibliography}

\end{document}